\documentclass{aa501}

\input epsf

\begin{document}

\title{ISOPHOT - Photometric Calibration of Point Sources}

\author{B.~Schulz 	      \inst{1,} \inst{2}
\and	S.~Huth 	      \inst{2,} \inst{3}
\and	R.J.~Laureijs 	      \inst{1,} \inst{2}
\and	J.A.~Acosta-Pulido    \inst{2,} \inst{3,} \inst{4}
\and	M.~Braun 	      \inst{2,} \inst{3,} \inst{7}
\and	H.O.~Casta\~neda      \inst{2,} \inst{3,} \inst{4}
\and	M.~Cohen 	      \inst{10,} \inst{11}
\and	L.~Cornwall 	      \inst{2,} \inst{5}
\and	C.~Gabriel	      \inst{1,}\inst{2}
\and	P.~Hammersley	      \inst{4}
\and	I.~Heinrichsen	      \inst{2,} \inst{6,} \inst{8}
\and	U.~Klaas 	      \inst{2,} \inst{3}
\and	D.~Lemke 	      \inst{3}
\and	T.~M\"uller 	      \inst{1,} \inst{2,} \inst{3}
\and	D.~Osip 	      \inst{9,} \inst{12}
\and	P.~Rom\'an-Fern\'andez \inst{1},\inst{2}
\and	C.~Telesco 	      \inst{9}
       }

\offprints{Bernhard Schulz, \email{bschulz@iso.vilspa.esa.es}}

\institute{
   ISO Data Centre, Astrophysics Division of ESA,
   Villafranca, P.O. Box 50727, 28080 Madrid, Spain
\and
   ISO Science Operations Centre, Astrophysics Division of ESA,
   Villafranca, P.O. Box 50727, 28080 Madrid, Spain
\and
   Max-Planck-Institut f\"ur Astronomie,
   K\"onigstuhl 17, 69117 Heidelberg, Germany
\and
   Instituto de Astrofisica de Canarias,
   38200 La Laguna, S/C Tenerife, Spain
\and
   Rutherford Appleton Laboratory,
   Chilton, Didcot, OX11 0QX, UK
\and
   Max-Planck-Institut f\"ur Kernphysik,
   Saupfercheckweg 1, 69117 Heidelberg, Germany
\and
   Astrophysikalisches Institut Potsdam,
   An der Sternwarte 16, 14482 Potsdam, Germany
\and
   Infrared Processing and Analysis Center,
   California Institute of Technology,
   MS 100/22, Pasadena, CA 91125, USA
\and
   211 Bryant Space Science Center,
   P.O. Box 112055, Dpt. of Astronomy,
   Univ. of Florida, Gainesville, FL 32611-2055, USA
\and
   Radio Astronomy Laboratory, 601 Campbell Hall,
   University of California, Berkeley, CA 94720, USA
\and
   Vanguard Research, Inc. Suite 204,
   5321 Scotts Valley Drive, Scotts Valley, CA 95066
\and
   MIT, Dept. of Earth, Atmospheric and Planetary 
   Sciences, Bldg. 54-420, 77 Massachusetts Ave., Cambridge MA, 02139
   }

\date{Received / Accepted}

\abstract{
All observations by the aperture photometer (PHT-P) and the 
far-infrared (FIR) camera section (PHT-C) of ISOPHOT included 
reference measurements against stable internal fine
calibration sources (FCS) to
correct for temporal drifts in detector responsivities. The FCSs
were absolutely calibrated in-orbit against stars, asteroids and
planets, covering wavelengths from 3.2 to 240~$\mu$m. We present
the calibration concept for point sources within a flux-range from
60~mJy up to 4500~Jy for staring and raster observations in
standard configurations and discuss the requisite measurements and
the uncertainties involved. In this process we correct for
instrumental effects like nonlinearities, signal transients, time
variable dark current, misalignments and diffraction effects. A
set of formulae is developed that describes the calibration from
signal level to flux densities. The scatter of 10 to 20\,\% of the 
individual data points around the derived calibration relations is
a measure of the consistency and typical accuracy of the
calibration. The reproducibility over longer periods of time
is better than 10\,\%. The calibration tables and algorithms have
been implemented in the final versions of the software for
offline processing and interactive analysis.
}

\maketitle


\section{Introduction}
In its broadest meaning, we define calibration as transforming
the specific, nowadays digital, output of a scientific instrument
to physical units. In our case these are flux densities
at given wavelengths and sky positions. This transformation
generally varies with every change of the instrument set-up,
hence the complexity of the calibration task increases with the 
number of instrument configurations used.
Initially it is derived on the grounds of the 
known instrument geometry and the relevant optical and
electrical properties, that as a whole we will refer to as the
ideal instrument model. Subsequently this model is refined and 
becomes more empirical, in order to match the measured data.

The task of calibration can be divided into three parts: First,
development of the instrument model and determination of the
instrument parameters that are assumed to be unchanging and 
that can be measured in the laboratory, e.g.
filter transmissions, aperture diameters, etc. Second, the
determination of the open parameters of the ideal
instrument model that can be determined only in-situ, i.e. with
the instrument built into the satellite and in the real
space-environment. And third, the determination of deviations
from the ideal instrument model that cannot be removed by
adjusting parameters, and require a new functional
property. These modifications to the instrument model are found
empirically, and generally originate in simplifications. 
Because open parameters and non-ideal 
instrumental effects are usually intertwined, 
an iterative process is required to separate and
quantify all the contributing effects.

This paper presents the photometric calibration of staring mode- 
and simple raster mode observations\footnote{Excluding the AOT P32.} 
of point sources with
the P- and C-sections of ISOPHOT (Lemke et al. \cite{lemke96}),
which is one of four scientific instruments on board ESA's
Infrared (IR) Space Observatory ISO (Kessler et al. \cite{kessler96}).
Unlike CCD devices, the detectors for the 
Mid-IR (MIR) and Far-IR (FIR) are far less stable
and exhibit a continuously changing relation between signal and incident
flux. Thus the two Fine Calibration Sources (FCS) built into ISOPHOT
played a crucial role as stable references for the photometry and most
of this paper will describe their empirical calibration.

The data collected for this task represent the largest part of all 
specific calibration observations during the mission. 
This resulted in a fairly homogeneous block of data.
Its analysis and comparison to modeled spectral energy distributions
(SEDs), of the observed celestial standard sources
yielded most of the results presented here and drove several 
refinements to the ideal instrument model.
An additional difficulty was the large number of possible instrument
configurations, which was limited somewhat by considering only one
standard aperture for each filter band of the aperture photometer
(see Table~\ref{wavtab}).
The set of instrument configurations and modes we treat herein
defines a well-understood baseline within the large parameter space,
where absolute calibration errors are expected to be 
minimal.
The calibration of further configurations and modes, like
chopped observations, extended source photometry, or non-standard 
apertures, is left to future publications.

We start with a brief review of the instrumental design 
with some emphasis on the internal reference sources
(Sect.~\ref{instrdes}), followed by an outline of the
calibration strategy (Sect.~\ref{calstrat}). Sect.~(\ref{sigcond}) 
continues with a description of the corrections
applicable to the detector signal. Sect.~(\ref{celcal}) presents
the celestial calibrators and Sect.~(\ref{phtcor}) describes 
those corrections imposed by photometric constraints. 
We derive the FCS calibration tables in
Sect.~(\ref{fcspow}) and present the final flux calibration,
with its mathematical description and a discussion on accuracy
and reproducibility, in Sect.~(\ref{flxcal}). A summary constitutes
Sect.~(\ref{summar}).

\section{Instrument Design} \label{instrdes}
\subsection{Optical System}
\begin{figure}
  \begin{center}
  \vspace{0cm}
  \hspace{0cm}\epsfxsize=5.8cm \epsfbox{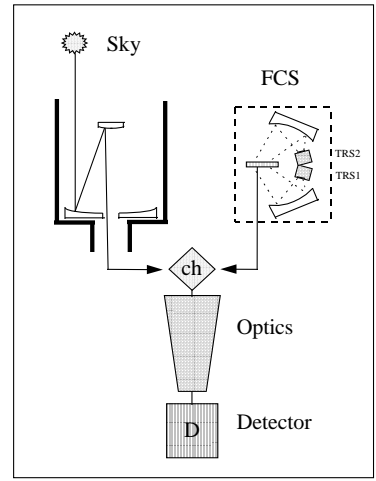}
  \caption{\it Instrument schematic from a calibrator's point of 
  view. The detector compares IR radiation directly from the 
  sky and from the internal reference source. Stability of the
  detector is required only for the time interval of the two 
  measurements.}
  \label{fcsgra}
  \end{center}
\end{figure}

A schematic instrumental set-up is shown in Fig.~\ref{fcsgra}. A
more detailed description can be found in Klaas et al.
(\cite{klaas94}) or Laureijs et al. (\cite{laureijs2001}). 
The set-up allows a choice between two sources of 
IR radiation via a chopper mirror:
i)~the telescope and 
ii)~the internal FCS.
The radiation of either source is analysed by an optical system
through filters, apertures, etc., and reaches finally
the IR detectors.

The telescope is a Ritchey-Chr\`etien design, diffraction-limited
at 5~$\mu$m, with an entrance aperture 60~cm in
diameter and an $f$-ratio of $15$ (Kessler et al.
\cite{kessler96}). Radiation reaches the chopper mirror
within ISOPHOT via a pyramidal mirror, which is centred
between the 4 ISO-instruments.
The sky is projected onto the focal plane with a scale of
0.04363~mm/$\arcsec$.

\subsection{Detectors} \label{respsect}
The subsystems PHT-P and PHT-C are a multi-aperture photometer 
with 3 single pixel detectors and a FIR camera comprising 2
detector arrays of $3 \times 3$ and $2 \times 2$ pixels.
The detectors of PHT-P are P1 (3.15-17.5~$\mu$m), P2 (15-30~$\mu$m), 
and P3 (40-130~$\mu$m) with 10, 2, and 2 filter bands respectively.
Each detector can be combined freely with one of 11 circular field 
apertures with diameters between 5\arcsec and 180\arcsec. 
Only one filter band is measured at a time.
The PHT-C pixel sizes are 1.9$\times$1.9~mm and 3.9$\times$3.9~mm with 
gaps of 0.1~mm between pixels. The projected pixel sizes on the sky are 
43.5\arcsec$\times$43.5$\arcsec$ for C100 and 
89.4\arcsec$\times$89.4$\arcsec$ for C200.

The detectors, made from doped silicon or doped germanium, are extrinsic 
photoconductors operated at temperatures of 1.7 to 3.6~K, depending
on material. 
Their response to IR radiation is characterised by the detector
responsivity, $R$, which is the parameter most relevant 
to calibration. It is given by the
relation between detector-signal and incident in-band power
$ R=\frac{(S - S_\mathrm{dark}) \, C_\mathrm{int}}{P} $, 
where $S$ is the detector signal expressed in
V/s for an integrating amplifier. $S_\mathrm{dark}$ is the relaxed
signal under dark conditions, $C_\mathrm{int}$ is the integrating
capacity (140 fF for C200, 90 fF for others) and $P$ represents 
the in-band power in Watts. 

The responsivity of these detectors is not constant. 
Depending on material, variations of a factor of 3 during
one orbit were observed. Changes of $R$ are induced by the
ionizing radiation in space and by the previous sequence of 
IR fluxes to which the detector has been exposed. We
consider $R$ to be stable over typical time intervals 
of around 10 to 20~min. To bring $R$ back to a nominal value, 
the detectors routinely underwent a curing-procedure after
having crossed the Radiation Belts during perigee and before
regular observations started (``science window") (see Laureijs
et al. \cite{idum}, \cite{laureijs2001}). 
These procedures consisted of detector specific combinations 
of detector-heating, bias increase or illuminator flashes 
(Lemke at al. \cite{lemke96}) and brought the responsivity back
to within 5\,\% of its nominal value. 
The doped-Ge detectors, which showed the biggest responsivity 
changes, underwent a second curing procedure after 8~hours, 
close to apogee.

\subsection{Fine Calibration Source (FCS)}
The instrument hosts two FCSs, of which FCS\,1 was used
for the staring and raster observations and, hence, is chiefly treated 
in this paper.
Both devices were manufactured at MPIA Heidelberg.
Each FCS contains two thermal radiation sources (TRS) made of
small diamond-plates of $1 \times 1 \times 0.1$~mm, coated with
Ni-Cr. The plates are heated electrically with stabilised
heating powers up to 50~mW, providing highly
reproducible IR fluxes. The stabilisation circuit works digitally, 
with a resolution of 12.21~$\mu$W. This assures a flux stability of better
than 4\,\% for heating powers above $h=120~\mu$W, and 2\,\%
above $h=264~\mu$W.

The radiation of both sources is combined in a diamond beam-splitter. 
The beam of TRS\,1 is reflected with practically no losses, while the 
beam of TRS\,2 is transmitted through the material and attenuated
by a factor of $\approx$1000. Only one TRS is heated at a time. The
attenuated TRS\,2 is operated with the doped-Ge detectors P3, C100 and
C200, which show higher responsivities, whereas the doped-Si
detectors P1 and P2 are used with TRS\,1 in the stronger 
reflected beam. Mirrors focus the emerging radiation in such a
way that a beam with the same aspect ratio as the telescope is
emitted, while the TRSs are imaged onto the chopper mirror, 
making the FCS appear to the detectors as a homogeneously
illuminated source. 

The in-band power provided by the FCS could be adjusted by changing
its heating power.
The fundamental calibration curves, in-band power versus FCS heating 
power, were initially determined by a grey body model 
(Schulz \cite{schulz93}), fitted to a few already known data points. 
The grey body is defined as $\epsilon B_\nu(T)$, which is the
Planck function multiplied by a wavelength-independent emissivity
$\epsilon$. The temperature, $T$, is linked to the FCS heating power,
$h$, by an empirical function, 
$T(h)=\alpha * P^\beta + \gamma + \delta \, \mathrm{atan}(P/2)$, where 
$\gamma$ is the temperature of the optical support structure (2.76~K), 
and $\alpha$, $\beta$ and $\delta$ are dimensionless constants. 
Typical values for these constants of FCS\,1/TRS\,2 are 
$25.8\pm 1.8$, $0.518\pm 0.006$ and $0.07\pm 0.04$ respectively.
An attenuation factor which modifies the
effective solid angle individually for each detector subsystem
was introduced, describing flux losses that were not predicted
by the grey body model.

The advantage of the grey body model was to allow predictions 
for other filter bands, when only a few data points
per detector were available. To increase accuracy, once enough 
data points were available, 
the covered ranges were interpolated by smooth low-order polynomials 
and the model was used for extrapolations only.

\section{Calibration Strategy \label{calstrat}}
Given the drifting detector responsivity, the calibration had to
be reestablished periodically by observing a known reference source.
To avoid time-consuming slews across the sky to celestial 
calibration sources, we used the FCSs as an intermediate reference
that could easily be compared with the telescope beam 
(Fig.~\ref{fcsgra}). 
All ISOPHOT observations that were implemented as Astronomical 
Observing Templates (AOT)(see Klaas et al. \cite{klaas94}), were 
designed to contain at least one FCS measurement.

The absolute calibration of these secondary standards was 
established in-orbit by comparing their output to known celestial 
standards, an activity that started during the performance verification
phase (PV), but continued throughout the mission due to the visibility 
constraints of some of the sources.
To minimize errors due to potential detector nonlinearities, the
FCS heating power was adjusted so that the emitted FCS flux
roughly matched the flux emitted by celestial source and background.
The limited linearity of the system demanded celestial calibration 
sources at all flux levels in all 25 filter bands.

\section{Deriving the Detector Signal} \label{sigcond}
To calculate the detector responsivity, $R$, we need to determine the detector signal. 
Ideally the signal is proportional to the detector current,
exhibits only Gaussian distributed noise, and remains constant over
periods of constant detector illumination.
To approach this ideal in practice, a number of effects must 
be treated, which lead to the signal conditioning 
procedures described below.

\subsection{Integration Ramps}

Integrating cold ($\approx$~3~K) readout electronics (CRE) 
(Dierickx et al. \cite{dierickx89}) are used to amplify the
currents of typically $10^{-16}$\dots$10^{-13}$~A that flow
through the ISOPHOT photoconductors (Lemke et al. 
\cite{lemke96}). The photocurrent is measured from the rate at which the
voltage at the charge capacitor within the CRE increases with
time. The voltage is sampled at regular time intervals for a
given duration before it is reset (non-destructive readouts, NDR,
and destructive readouts, DR). The readouts of all channels of 
a CRE unit are time-multiplexed and sent via a single line to
the external electronics unit (EEU) outside the cryostat. 
There they are further amplified and digitized by a 12-bit 
analog-to-digital-converter (ADC). The sequence of samples
between two resets is called an integration ramp and ideally fits a
straight line. In the following we will refer to the slope of
the fitted line as ``signal'', measured in V/s. Multiplication
by the integration capacity, for which the design value of 90~fF
(for C200 140~fF) is adopted, leads to the photocurrent in Amps.

\subsection{Nonlinearities of Integration Ramps}

\begin{figure}
  \vspace{0cm}
  \hspace{0cm}\epsfxsize=88mm \epsfbox{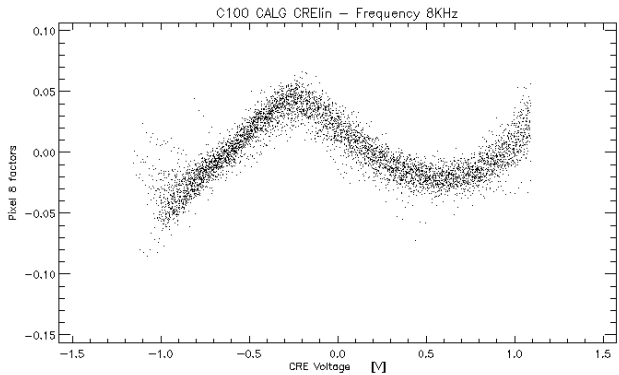}
 \caption{\it Ramp nonlinearity for pixel 8 of C100. Deviations of
 many integration ramps from a straight line are plotted against
 CRE output voltage. This relation is used for
 ramp-linearisation.}
  \label{crelin}
\end{figure}

The actual integration ramps are, however, not perfectly straight
(see Fig.~\ref{crelin})(Schulz \cite{schulz93}).
Two effects can be separated: 
i) The AC-coupled CRE circuit is not a perfect integrator. The 
output voltage can rise only to the level of the detector bias
times the small amplification of the circuit ($\approx$10). This
saturation level is reached asymptotically via a typical
RC-loading curve. High bias voltages above 10~V have
sufficiently high saturation levels ($\approx$100~V and more) 
that they show practically linear integration ramps within 
the CRE output range of $\approx$2~V. However, for the smaller
biases of P3, C100 and C200, the saturation voltage drops such
that the curvature of the loading curve appears within the
dynamic range of the integration ramp (debiasing).
ii) Secondly, all integration ramps show deviations from a 
straight line that always appear at the same CRE output voltage,
regardless of the level from which the ramp started to integrate. Due to
the direct link to the CRE output voltage we attribute this 
component to an intrinsic nonlinearity of the amplifier 
within the circuit.

The correction algorithm is based on the observation that all
integration ramps of one detector pixel can be matched by
conserving the CRE output voltage of the individual readouts,
while linearly stretching and shifting the time-axis of each
ramp (Schulz \cite{schulz93}). In this way an average 
integration ramp was determined for each detector pixel from
a number of measurements with ramps of large dynamic range.
The difference between a straight-line fit to this ramp and the 
ramp itself (Fig.~\ref{crelin}) is used to linearise 
individual ramps. The maximum absolute values range between 
40 and 100~mV over the maximum CRE dynamic range of 2~V.
The correction depends only on CRE output voltage, and has no
further dependences on readout frequency, in-band power or time. 

The reliability of the correction is deduced from the standard
deviation of the individual ramps from the average
integration ramp. Over most of the dynamic range it is below
$\pm10$~mV, with a tendency towards larger scatter for longer
wavelength detectors. The highest standard deviation appears in
the central pixel of C100, showing $\pm20$~mV. The higher
scatter towards long wavelength detectors with smaller biases
is interpreted as resulting from a simplification in the
correction algorithm. Relating the corrections only to the
CRE output voltage neglects the fact that the curvature due to
debiasing also depends on the reset level of the integration
ramp, which shifts slightly w.r.t. the CRE output voltage. Hence
an additional scatter appears, increasing with smaller bias.
Considering the maximum standard deviation of $\pm20$~mV and an
average dynamic range of 1 V (half the maximum dynamic range), 
we estimate the typical residual systematic error to be about 
$\pm$2\,\% after correction.

\subsection{Ionizing Radiation Effects and Deglitching}

\begin{figure}
  \vspace{0cm}
  \hspace{0cm}\epsfxsize=88mm \epsfbox{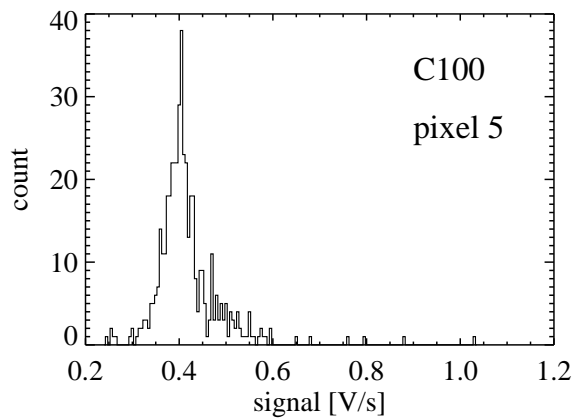}
 \caption{\it Histogram showing the signal distribution in a 
 long measurement of the central pixel of C100. The tail
 towards higher signals is caused by high energy particle hits
 and can be removed by appropriate algorithms.}
  \label{sigdistr}
\end{figure}

The output signal is disturbed by energetic particle hits 
(glitches), mostly protons and electrons. A typical
distribution of fitted slopes after ramp linearisation is shown
in Fig.~\ref{sigdistr}. The asymmetry of the distribution is
caused by glitches that instantaneously increase the charge
on the capacitor during the integration, so that the average
slope of the affected integration ramp is increased. Typically,
the readouts following the glitch continue integrating as 
before; however, stronger hits can affect the detector, 
leading to long-lasting responsivity changes. A number of 
algorithms have been developed to remove 
signals affected by glitches (Gabriel et al. \cite{gabriel97}).

\subsection{Digitization Effects}

All deglitching algorithms still
leave an asymmetry in the signal distribution.
For stronger asymmetries the simple average is
often quite far from the peak of the distribution.
The median comes closer to the peak; however, it is a bad choice
for very weak signals and short integration ramps, where
the bins of the A/D-conversion become significant. The 
AC-converter covers the range from $-10$ to $+9.995$~V. Dividing
by the gain factor of the analog electronics and $2^{12}$ for
12 bits, the least significant bit is equivalent to $6.1$~mV. For integration
ramps of $1$~second this would yield a detector current of
$5.5\times10^{-16}$~A. As an example, the $100$-$\mu$m channel
would produce discrete slope levels in about 20~mJy intervals.
Since the result of the median is always an existing value of the 
sample, the accuracy is limited to the separation of the intervals.
In this case, better results can be obtained by fitting a Gaussian
to the signal distribution.

\subsection{Detector Transients and Long-Term Drifts}
After a change of IR flux, the photocurrent shows a
characteristic behaviour before it approaches the final value
asymptotically, which we call hereafter `transient'. 
Relaxation times of several minutes are common.
The second effect, which we refer to as drift, is also
observed as a gradual variation of the responsivity; however, it is on
timescales of hours rather than minutes. Drift is due to the
continuous bombardment of the detector material by high energy
particles in space (e.g. see Blum et al. \cite{blum90}), 
but can also be triggered by strong flux changes.
In the following we describe in more detail how the
transient correction was applied to the FCS calibration data,
since this was an important step in eventually achieving
consistent photometric results.

\subsubsection{Solutions for Transients}
The observed detector current is the sum of dark current and a 
current proportional to the number of incident IR photons (Bratt
\cite{bratt77}; Sclar \cite{sclar84}). However, this is 
correct only for a system that has reached equilibrium after each
flux change. To derive signals that are independent of
measurement time, we need to determine the signal as $t
\rightarrow \infty$.
The simplest solution is to determine the level when the
signal has practically reached its final value and no trend is
detected within the noise. 
To filter out the cases that need special treatment, a 
`Stability Check', implemented in the PHT Interactive Analysis 
(PIA) software (Gabriel et al. \cite{gabriel97}), is conducted. The
method tests successively smaller pieces of data towards the end of a 
measurement for stability at a 95\,\% confidence level. The
smallest piece is limited to either 7 data points or 8~sec.
Stable signal levels were detected for P1, P2 and C100 for 
about 65\,\% of our calibration measurements. Typical measurement 
times were between 64 and 256~sec. For fainter signals, however, the
relaxation times generally increase and other methods 
had to be applied to predict the final signal level from 
only the first part of the transient.

\subsubsection{Empirical Transient Fit}
\begin{figure}[h!]
  \vspace{0cm}
  \hspace{0cm}\epsfxsize=88mm \epsfbox{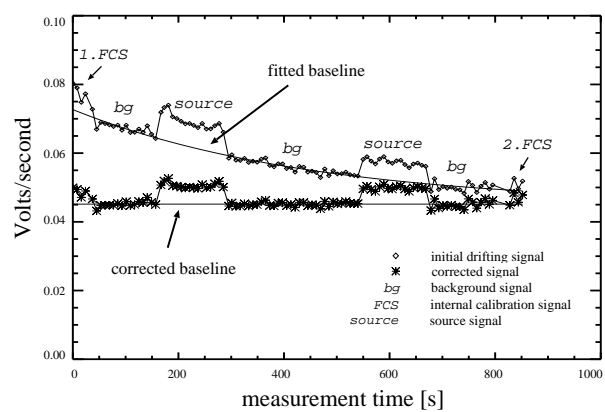}
 \caption{\it Observations in `nodding mode' with a linear
  3-step raster used for faint sources. Several changes of the 
  pointing between a strong background and a faint source on top
  of it make the source signal appear as a modulation on top of
  a strong transient. This is corrected by dividing by a 
  baseline fitted to all background levels. The sequence starts 
  and ends with FCS measurements.}
  \label{fstab01}
\end{figure}

Transients converge asymptotically towards a final value,
$S_{\infty}$. Empirical functions that fit this characteristic
are usually constructed from exponentials. Since long term
detector drifts bound the integration time above, 
typical durations of `staring' measurements at one
flux level range between 64 and 256~sec. We use an empirical
function of the form 
$S(t)=S_{\infty}+(S_{0}-S_{\infty})\mathrm{e}^{-ct}$. $S_{0}$ is the
signal before the flux change, $t$ is time and $c$ is a free
parameter. Since in most cases $S_{0}$ is not known, it is
also left as a free parameter to fit.
Stronger transients that are `remembered' by the detector during
successive measurements can be eliminated by fitting a
`baseline' of the same expression, if more than one measurement 
of the same flux level is obtained (see Fig.~\ref{fstab01}).

\subsubsection{Automatic Validation of Transient Fits}
Automatic procedures
were developed to reduce the
calibration data homogeneously. The routines include stability test,
empirical transient fit and error 
discrimination\footnote{Note that the quality check 
function was specifically developed
to process FCS calibration measurements automatically. Although
this check is not available in the PIA software, the described
transient correction can be interactively applied by the user,
so that our conclusions are not necessarily limited to our 
specific set of measurements only.}
(Huth \& Schulz \cite{huth98}). 
If the stability test cannot locate
a stable part of the measurement within the criteria given above, 
a transient fit is initiated.
The criteria that were developed during intensive tests to 
eliminate dubious results from transient fitting are as follows:

(i) We calculate the average of the last 30\,\% of the measurement
and call it $S_{30}$. The valid fit result must deviate by no 
more than $\pm50$\,\% from $S_{30}$.

(ii) A straight line is fitted to the last 40\,\% of the
measurement and extrapolated to a point at 2 times the
measurement time. 80\,\% of the difference between this point and
$S_{30}$ is the maximum difference allowed between the signal
level predicted by the transient fit and $S_{30}$.

(iii) For each fitted transient, the term 
$(1-S_{1}/S_{2})/(t_{1}-t_{2})$ is evaluated twice, where 
S and t are the slopes and times of either the first two, 
or the last two, data points in the measurement.
Absolute differences between these two results that are smaller 
than 0.001~s$^{-1}$ are rejected. This and 
criterion (ii) buffer the curvature of the fit against small
values since small curvatures result in very uncertain
asymptotes.

In case these criteria fail, we take $S_{30}$ as the best guess
for $S_{\infty}$ but the result is flagged as being less reliable. 
This applied to about half the cases where transient fits 
were attempted.
We caution that the criteria were tuned empirically to this
large, but special, sample of FCS calibration measurements. 
We assume, therefore, that they work best for measurement
times $\ge 64$~s.

\subsubsection{Signal Uncertainties}
In the automatic processing we distinguish three types of
measurement.
The first consists of measurements that reach stability
within their integration time. Here the statistical error is the
standard deviation of the mean of the stable part of the
measurement. The second type consists of
transient fitted measurements that meet the criteria above, 
and we use $\sqrt{1/(n-1)\sum_{i}^{n}(S^\mathrm{modl}_{i}-S^\mathrm{meas}_{i})}$ as 
error, where n is the number of valid slopes in the measurement,
and $S^\mathrm{modl}$ and $S^\mathrm{meas}$ are the modeled and 
measured slopes respectively. 
If the transient function were a
constant, i.e. a horizontal line, the expression would be
the standard deviation. We find that using the above equation, 
instead of the standard
deviation of the mean, provides more realistic errors. Finally,
in cases where no stability is found and the transient fit
fails, the error assigned is the standard deviation of the last
30\,\% of the measurement. Thus the errors of the three types
also weight the results with respect to their 
systematic uncertainties.

\subsection{Signal Dependence on Readout Timing}
\begin{figure}
  \vspace{0cm}
  \hspace{0cm}\epsfxsize=88mm \epsfbox{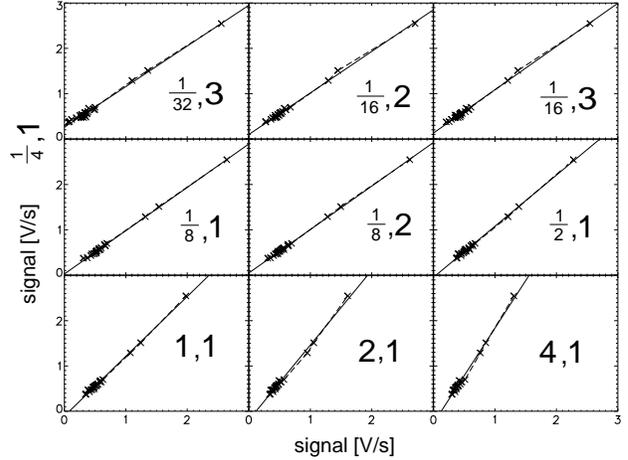}
 \caption{\it The diagram shows the relations of signals of the
 same in-band power measured with different readout timing settings. 
 The signals on the y-axis are measured using a 1/4~s reset
 interval and a data reduction factor of 1. The reset interval
 and data reduction factor, respectively, for the signals of the x-axis are
 given by the two numbers in each diagram. The 
 data points follow a linear relation, with slope different from
 unity and non-zero x-intercept. This example was measured for
 detector C100.}
  \label{crelin1}
\end{figure}

The readout timing (ROT) of the CRE was selected 
individually for each measurement according to 
the observer's flux estimate, to give a maximum dynamic range
without saturation and thus a minimum of readout and 
sampling-noise. It is controlled by 3 parameters: the time
interval between two destructive readouts, $t_\mathrm{reset}$; the
number of non-destructive readouts, $n_\mathrm{NDR}$; and the data
reduction factor. For the range of $128$~s to
$1/64$~s for $t_\mathrm{reset}$, depending on the detector, a total 
of 14 to 15 combinations was actually used during operations.

We observed several non-variable\footnote{For the time of the 
observation.} celestial sources (stars, planets) with all
ROT combinations suitable for in-band powers that cover a range of a 
factor of 16 above and below that actually observed. Care was
taken to ensure the stability of the detector response during
the measurement series by repeating the set of ROT combinations
in reverse order. Ideally one expects to find the same signal
within the errors for the same source, regardless of CRE set-up.
However, even after correcting for ramp nonlinearity, we
derived different signals for the same source. The signals show
a strong dependence on the reset interval, especially for those
cases where part of the integration ramp exceeds the
dynamic range and is discarded due to saturation.

To achieve consistency, we correct all signals so
they appear as if measured with a common
reset interval of $t_\mathrm{reset}=1/4$~s and a data reduction factor
of 1. The signal, $S$, is corrected according to $S^{\prime} = A0 +
A1 \times S$, where the parameters $A0$ and $A1$ depend on the
ROT set-up. They were determined from a least-squares fit to the
calibration data. An example is shown in Fig.~\ref{crelin1}. It
should be noted that $A0$ is generally different from zero and
the slope $A1$ is significantly different from unity.
All pixels of equal detector-subsystems were found to follow the
same relation, suggesting that a pure CRE-effect is observed.
Again the residual errors of the correction are systematic in 
nature. The scatter of the calibration data around the fits
suggests typical systematic errors of no more than 5\,\%
and probably better than 2\,\%.

\subsection{Dark Current}
\label{darksect}
\begin{figure*}
  \begin{center}
  \vspace{0cm}
  \hspace{0cm}\epsfxsize=180mm \epsfbox{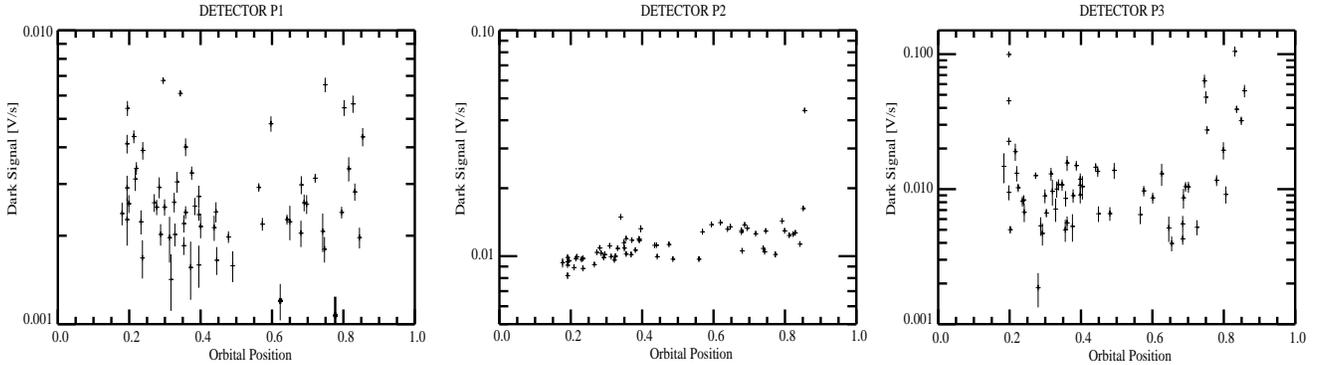}
  \caption{\it The evolution of the dark signal with relative
  orbital position. The data points were taken
  randomly during the mission and plotted versus orbital position,
  where the range 0 to 1 corresponds to the full 24 hour orbit,
  starting at perigee.
  }
  \label{P_dark}
  \end{center}
\end{figure*}
\begin{figure*}
  \begin{center}
  \vspace{0cm}
  \hspace{0cm}\epsfxsize=180mm \epsfbox{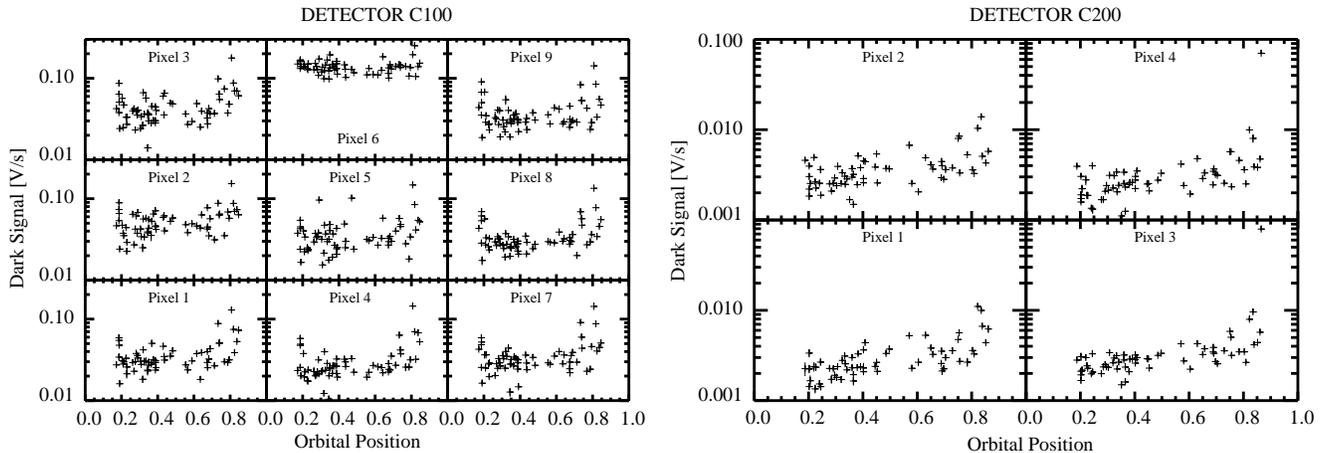}
  \caption{\it As for Fig.~\ref{P_dark} but for the detectors 
  C100 and C200.}
  \label{C_dark}
  \end{center}
\end{figure*}
Even under completely dark conditions and after relaxation of
transients due to previous illumination, a
detector current is measured, arising from thermal generation
of free charge carriers and surface leakage
conductance (Bratt \cite{bratt77}).
This value must be subtracted to obtain the pure photocurrent.
The dark current had already been determined for each 
detector pixel in the laboratory; however, in orbit 
higher dark currents than on the ground were generally observed. 
A comparison is given by Lemke et al. (\cite{lemke96}).
We interpret part of the increase as due to the faint end 
of the glitch distribution, where many small glitches produce a 
continuous current (see Fig.~\ref{sigdistr}).
Dark signals measured randomly over the mission versus relative 
orbital position are plotted in Figs~\ref{P_dark} and 
\ref{C_dark}. The range 0 to 1 on the x-axis corresponds to the
full 24-hour revolution, starting at perigee. The data show a
dependence of the dark signal on orbital position. The
systematic rise along the orbit is attributed to 
high energy particles in space. In particular, the steep
rise at the end indicates the beginning of entry into the
Radiation Belts. The uncertainty in the dark
signals is due to detector noise and limitations in
correcting detector transients. A low order polynomial fit to 
these diagrams is used to model the time-dependence of the
dark current over the orbit.

\section{Celestial Calibrators} \label{celcal}
The absolute photometric calibration of ISOPHOT is based entirely 
on point sources, which constitute the best known IR flux densities
in the sky. The sources were observed either for PHT-P at the
centre of a standard circular aperture that is defined per 
filter\footnote{The standard apertures were chosen to include
at least 90~\% of the energy of the point-spread-function (PSF).}
(see Table~\ref{wavtab}) or, for the PHT-C array
detectors, at the centre of each individual pixel. As the
detailed radiation spectrum of the FCS was unknown and
linearity of the detectors could not a priori be postulated,
the empirical FCS calibration required the availability of
calibration standards covering the full observable flux range
between 3 and 240~$\mu$m. The upper limit of the absolute
flux range is determined by the dynamic range of the
CRE and the detector responsivity.
The lowest signals are mainly determined by the celestial 
background and the detector responsivity. 

\begin{figure}[h!]
  \vspace{0cm}
  \hspace{0cm}\epsfxsize=88mm \epsfbox{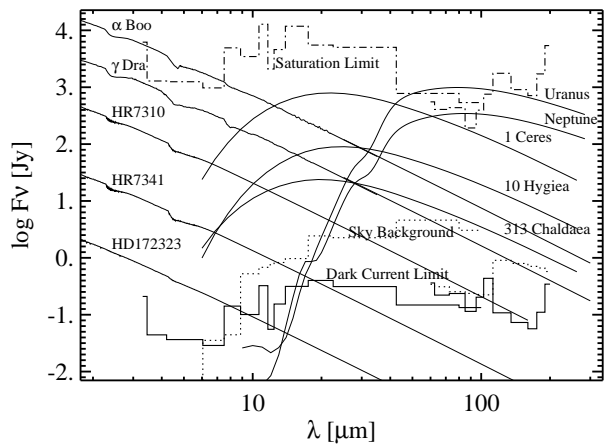 \epsfclipon}
  \caption{\it The coverage of the flux domain accessible to 
  ISOPHOT by the three kinds of celestial standard.}
  \label{flxreg}
\end{figure}

In Fig.~\ref{flxreg} we illustrate the
coverage of the flux domain accessible to ISOPHOT by celestial
standard sources.
The largest background components are zodiacal thermal re-emission, peaking
around 20~$\mu$m, and Galactic cirrus, contributing
most at the longest wavelengths. Only at the shortest 
wavelengths does the dark current dictate the lowest signals.
The dotted line shows typical surface brightnesses for 
celestial backgrounds, converted to an equivalent flux density
of a point source, observed in the standard aperture without
background. The upper and lower absolute flux limits are
plotted as dashed lines. Note that the
limiting sky background for the overlapping wavelengths of
PHT-P and PHT-C is different due to their different aperture
areas. Examples of each of the three types of
calibration source (planets, asteroids, stars) are plotted as 
solid lines.

\subsection{Stars}
\begin{table*}
\begin{center}
\footnotesize
\begin{tabular}{lcccccccccc} 
\multicolumn{8}{c}{\bf Stellar Celestial Standards}                                 &\multicolumn{3}{c}{\bf Solar System Standards}                      \\
\hline\noalign{\smallskip}
Name        & Common            &   Src. &   Spec.   & F$_\nu^{12\mu{\rm m}}$&(P\_11.5) &
                                                    F$_{\nu}^{60\mu{\rm m}}$&(C\_60)   &       Name                  &F$_{\nu}^{60\mu{\rm m}}$&   (C\_60)      \\
	
            &  Name          &        &  Type  & Jy   &   mag  & Jy   &  mag  &                                    &    Jy        	 &     mag        \\
\hline\noalign{\smallskip}
\object{HD172323}    & -              &    PH  &  F9V	   & 0.065  &  6.55  &	     &	     &    \object{Uranus}  	  & 913 \dots  972	 & -7.28\dots-7.21\\
\object{HD184400}    & -              &    PH  &  F5	   & 0.027  &  7.53  &	     &	     &    \object{Neptune} 	  & 301 \dots  317	 & -6.07\dots-6.01\\
\object{HR337}$^*$   & $\beta$ And    &    MC  &  M0IIIv   & 192    & -2.05  & 7.97  & -2.17 &	  \object{1 Ceres}        & 248  \dots 433	 & -6.46\dots-5.85\\
\object{HR617}$^*$   & $\alpha$ Ari   &    MC  &  K2III    & 55.7   & -0.76  & 2.33  & -0.83 &	  \object{2 Pallas}       & 55.3 \dots 208	 & -5.67\dots-4.23\\
\object{HR1457}$^*$  & $\alpha$ Tau   &    MC  &  K5III    &	    &	     & 19.1  & -3.12 &	  \object{3 Juno}         & 37.6 \dots 62.2	 & -4.36\dots-3.81\\
\object{HR1654}      & $\epsilon$ Lep &    MC  &  K5IIIv   &        &	     & 1.61  & -0.43 &	  \object{4 Vesta}        & 153  \dots 291	 & -6.03\dots-5.34\\
\object{HR3748}      & $\alpha$ Hya   &    MC  &  K3II-III & 93.4   & -1.31  & 3.92  & -1.39 &	  \object{10 Hygiea}      & 29.0 \dots 43.9	 & -3.97\dots-3.52\\
\object{HR5340}$^*$  & $\alpha$ Boo   &    MC  &  K1III    & 506    & -3.15  & 21.0  & -3.22 &	  \object{54 Alexandra}   & 20.0 \dots 26.1	 & -3.42\dots-3.13\\
\object{HR5886}      & -              &    PH  &  A2IV	   & 0.251  &  5.09  &	     &	     &	  \object{65 Cybele}      & 13.4 \dots 22.3	 & -3.24\dots-2.69\\
\object{HR5986}      & -              &    PH  &  F8IV-V   & 2.24   &  2.71  &	     &	     &	  \object{106 Dione}      & 3.80 \dots 3.93	 & -1.35\dots-1.32\\
\object{HR6514}      & -              &    PH  &  A4V	   & 0.095  &  6.14  &	     &	     &	  \object{313 Chaldaea}   & 7.74 \dots 8.14	 & -2.15\dots-2.10\\
\object{HR6688}      & -              &    PH  &  K2III    &        &	     & 0.463 &  0.92 &	  \object{532 Herculina}  & 16.3 \dots 44.5	 & -3.99\dots-2.91\\
\object{HR6705}$^*$  & $\gamma$ Dra   &    MC  &  K5III    & 108    & -1.46  & 4.55  & -1.56 &		  		  &			 &		  \\
\object{HR6847}      & -              &    MC  &  G2V	   & 0.337  &  4.76  &	     &	     &		  		  &			 &		  \\
\object{HR7001}$^*$  & $\alpha$ Lyr   &    MC  &  A0V	   & 27.0   &  0.00  &	     &	     &		  		  &			 &		  \\
\object{HR7310}      & -              &    MC  &  G9III    & 14.5   &  0.67  & 0.586 &  0.67 &		  		  &			 &		  \\
\object{HR7341}      & -              &    PH  &  K1III    & 0.951  &  3.64  &	     &	     &		  		  &			 &		  \\
\object{HR7451}      & -              &    PH  &  F7V	   & 0.427  &  4.51  &	     &	     &		  		  &			 &		  \\
\object{HR7469}      & -              &    PH  &  F4V	   & 1.07   &  3.51  &	     &	     &		  		  &			 &		  \\
\object{HR7633}      & -              &    PH  &  K5II-III & 9.66   &  1.14  & 0.391 &  1.11 &		  		  &			 &		  \\
\object{HR7742}      & -              &    PH  &  K5III    & 3.66   &  2.18  &	     &	     &		  		  &			 &		  \\
\object{HR7980}      & $\omega$ Cap   &    MC  &  M0III    & 	    &	     & 1.17  & -0.08 &		  		  &			 &		  \\
\object{HR8684}      & -              &    PH  &  G8III    & 8.45   &  1.27  &	     &	     &		  		  &			 &		  \\
\object{HR8775}$^*$  & $\beta$ Peg    &    MC  &  M2.5II-I & 265    & -2.44  & 11.2  & -2.54 &		  		  &			 &		  \\
\hline\noalign{\smallskip}
\end{tabular}
\caption{\it The list of stars, planets and asteroids used for ISOPHOT calibration. The
flux densities $F_{\nu}$ are given at the reference wavelengths of
the filters P\_11.5 and C\_60, depending if the respective standard
was used in the range 3-30\,$\mu$m or 45-240\,$\mu$m. In addition 
we list the magnitudes in the ISOPHOT photometric system with the Vega
model SED as zero point. 
The identifiers MC for M. Cohen and PH for P. Hammersley indicate the SED 
version that was actually used for calibration.
Asterisks indicate absolute validation by MSX (Cohen et al. \cite{cohen2001}).
For solar system objects we give the brightness range observable 
over the ISO mission.
}
\label{allsrctab}
\end{center}
\end{table*}
A large database of SEDs of standard stars was prepared in
support of the ISO calibration. Important aspects of the source
selection were non-variability, no known stellar companions, no 
IR excess, low sky background and good visibility by ISO
throughout the mission.

The ISO Ground Based Preparatory Programme (GBPP) 
(Jourdain \& Habing \cite{muizon92}; van der Bliek et al. \cite{bliek92}; 
 Hammersley et al. \cite{hammersley98}) provided SEDs
between 1 and 160~$\mu$m by interpolating in the Kurucz stellar model
grids (Kurucz \cite{kurucz93}), using temperature, surface gravity and
metallicity. The temperatures were derived either from the IR
Flux Method (Blackwell et al. \cite{blackwell91}, \cite{blackwell98}) 
or the V-K versus T$_\mathrm{eff}$
relationship (Di Benedetto \cite{benedetto93}, \cite{benedetto98}). 
Surface gravities and metallicities were from 
Cayrel de Strobel et al. (\cite{cayrel92}). The spectral
shapes derived from the grids were then normalized using 
near-IR (NIR) photometry (Kn or K-band). 
The attributed errors range between 3\,\% and 5\,\% which is
dominated by the error in conversion to an absolute flux density scale. The random
scatter between the measured and predicted fluxes at 10~$\mu$m is shown
to be better than 1.5\,\% (Hammersley et al. \cite{hammersley98}).

Another calibration programme by Cohen et al. (\cite{cohen96})
provided {\it empirical} SEDs in the range from 1.2 to 35~$\mu$m by splicing
together measured spectral fragments of cool K- and M-giants and
calibrating them by absolutely calibrated Kurucz models of
Sirius and Vega. The measured
``composite" SEDs are used to derive so-called ``template" SEDs for
fainter stars (Cohen et al. \cite{cohen99}), assuming that the intrinsic
spectral shape depends only on spectral type and 
luminosity class. The absolute flux level is set by
well-characterised NIR and MIR photometry, including IRAS data.
To extend those composite-SEDs not observed as far as 35~$\mu$m 
by the Kuiper Airborne Observatory, Engelke functions 
(Engelke \cite{engelke92}; Cohen et al. \cite{cohen95}, \cite{cohen96}), 
with effective temperatures from Blackwell et al.
(\cite{blackwell91}), were used. 
To create FIR extensions to support the longest wavelengths of 
ISOPHOT (240~$\mu$m), continuum model atmospheric spectra were 
attached to the empirical composites and extended to 
300~$\mu$m (Cohen at al. \cite{cohen96}).
The typical absolute accuracy
(i.e., in F$_{\lambda}$ or F$_{\nu}$ at any wavelength)
of the templates is about 3\,\%, while the FIR extensions
ranging from 25 to 300~$\mu$m have computed uncertainties of
about 6\,\%.

Note that both programmes deliberately use the 
{\it same} Vega model SED 
for their zero point definitions. Furthermore, the measured
zero points of the various photometric systems were determined 
to ensure that there would be no systematic differences 
depending upon which data were used (Cohen et al. \cite{cohen99}).
Specifically for ISO, both programmes supplied electronic 
versions of their SEDs with spectral resolutions of typically
50 to 300. The objects are listed in Table~\ref{allsrctab}.

The irradiances of a number of primary (Sirius, Vega),
secondary (bright K- and M-giants) and tertiary (template-SEDs)
calibration stars have been {\it absolutely} validated by a
dedicated radiometric calibration experiment carried out on the
US Midcourse Space Experiment (MSX: Mill et al. \cite{mill94}) 
by Cohen et al. (\cite{cohen2001}). The 1.2--35~$\mu$m 
spectra of seven of the stars in Table~\ref{allsrctab} have been validated 
by this means.
We note that because of its observed deviation from
the model SED longward of about 17~$\mu$m, we do not use observations of Vega 
to calibrate any long wavelength filter.

Brighter flux levels in the FIR (i.e. filters with reference
wavelengths $\ge$50~$\mu$m)) had to be calibrated by other objects,
since even the brightest standard star measured by ISOPHOT
($\alpha$ Boo) drops to less than 2~Jy at 200~$\mu$m (see
Fig.~\ref{flxreg}).

\subsection{Planets} 
\object{Uranus} and \object{Neptune} are commonly used as
submillimeter and FIR standards (Hildebrand et al.
\cite{hildebrand85}; Orton et al. \cite{orton86}; Griffin \&
Orton \cite{griffin93}). They provided the highest calibrated
flux levels that ISOPHOT could measure at long wavelengths.
Mars was already too bright when it became visible to ISO. 

In an effort to homogenize the calibration of ISOPHOT and the
Long Wavelength Spectrometer (LWS) in the long wavelength range, for the final version of 
the offline processing software (OLP 10) we changed the planet 
models used in earlier versions (Schulz et al. \cite{schulz99}). 
The ones supplied by Abbas (1997, priv. comm.) were replaced
by models produced by Griffin \& Orton for LWS.
The models are calibrated with 0.35 to 3.333~mm JCMT data 
(Griffin \& Orton \cite{griffin93}), which in turn were 
calibrated against Mars 
(Wright \cite{wright76}; Wright \& Odenwald \cite{wrightod80}).
At the short wavelength end, the model was constrained
by Voyager IRIS data from 25 to 50~$\mu$m by 
Hanel et al. (\cite{hanel86}) for \object{Uranus}, and
Conrath et al. (\cite{conr89}) for \object{Neptune}.
The temperature structure and composition of H$_{2}$, 
He and CH$_{4}$ was taken from Voyager radio occultation 
experiments by Lindal et al. (\cite{lindal87}) for \object{Uranus}, and
by Lindal (\cite{lindal92}) for \object{Neptune}.

We compared the models of Griffin \& Orton (GO) with 
the ones of Abbas (AB).
Above 45~$\mu$m the GO model of \object{Neptune} is $\approx$10\,\% 
brighter than AB.
In contrast, for \object{Uranus} the GO model fluxes are about 
3 to 10\,\% lower than AB above 50~$\mu$m
and deviate no more than 15\,\% above 20~$\mu$m.
At shorter wavelengths the discrepancies between AB and GO 
rise dramatically for both planets. However, these are no longer 
relevant to the PHT calibration, since the C\_50
filter passband starts only around 40~$\mu$m.
We take these differences as indicative of how well the fluxes
from \object{Uranus} and \object{Neptune} are known at FIR wavelengths and 
conclude that, for our purposes, they are still within a 
$\pm$10\,\% margin.

\subsection{Asteroids}
A small group of asteroids was chosen to be used as FIR calibration 
standards. They populate the intermediate-flux level gap 
that appears at wavelengths from 45 to 200~$\mu$m between \object{Uranus}, 
\object{Neptune} and the brightest stars (see Fig.~\ref{flxreg}).

Some specific observational complications arose, however, since 
they are moving objects with respect to the sky background and show
periodic variations of intensity due to rotation and
varying distance from both the Earth and Sun. 
Selection criteria were: well understood rotational behaviour, 
small lightcurve amplitude, some form of independent size 
determination (either direct imaging or via occultation measurements), 
good visibility during the ISO mission, and availability of sufficient 
observational data from visible to submillimetre wavelengths.
The selection resulted from a combination of
extensive ground-based observing campaigns at thermal wavelengths
from the IRTF, UKIRT, and JCMT, additional visible wavelength lightcurve
measurements, and a series of FIR observations from the now
retired KAO. These observations
were used to confirm the validity of the subsequent thermophysical 
modeling effort.

Initially, modified versions of the Standard Thermal Model 
(Lebofsky \cite{lebofsky89}) were used. For the final ISOPHOT
photometric calibration, a thermophysical model (TPM) assuming a
rotating ellipsoid and parameterising heat conduction, surface
roughness, and scattering in the regolith was adopted for 10
asteroids (M\"uller \& Lagerros \cite{mueller98}). 
The TPM is capable of producing thermal lightcurves and spectral
energy distributions for any time, taking into account 
the real observing and illumination geometry.
The overall comparison between model predictions and our large 
sample of MIR, FIR, and submillimetre observations 
(about 700 individual measurements between 2 and 2000~$\mu$m)
demonstrated an accuracy of approximately
$\pm$10\,\% across a wide wavelength range from 10-500~$\mu$m.
In a few cases, existing direct measurements of shape
significantly improved the model accuracies.

In a recent study of the accuracy of the TPM, ISOPHOT 
observations of asteroids were independently calibrated in the FIR
(M\"uller \& Lagerros \cite{mueller2001}, \cite{mueller2001a}).
In this work, parts of the FCS power curves were established
using measurements only of planets or stars. These were used
then to calibrate the asteroid observations and compare the 
results to the TPM predictions.
It was shown that observations and
predictions for \object{1 Ceres}, \object{2 Pallas} and 
\object{4 Vesta} agree within 5\,\%. For \object{3 Juno}, 
\object{10 Hygiea}, \object{54 Alexandra}, \object{65 Cybele}, 
and \object{532 Herculina} the agreement is still within 
10 to 15\,\%. 
For the 2 objects without independent ISO data the comparison 
to ground based thermal observations gave r.m.s. values of 
14\,\% for \object{313 Chaldaea} and 29\,\% for \object{106 Dione}.
Note that the r.m.s. values include also observational errors, 
which are dominated by the structured FIR sky background
and calibration uncertainties. Modeling limitations
are mainly due to uncertainties about the exact asteroidal 
shapes.
The final modeling results were
also confirmed to an accuracy of better than 10\,\% via comparison
with rotationally resolved target observations from the IRAS
database.
Table~\ref{allsrctab} lists the solar system objects
that were used for ISOPHOT calibration.

\subsection{In-Band Power Calculation}
To determine the responsivity of a detector we need to 
calculate the in-band power. Given the power distribution 
of a celestial standard in the focal plane, i.e. the 
normalized point spread function\footnote{The integral 
of the PSF over the focal plane is normalized to 1.}, 
$PSF(x,y,\lambda)$, times the flux density of the source,
$F_{\lambda}(\lambda) = (c/\lambda^2) F_{\nu}$,
we define the in-band power as the integral over the
product of i) the flux density of the source, ii) the PSF, 
iii) all spectral transmissions $T_{f}(\lambda)$
and iv) the relative spectral response function of the 
detector $R_{p}(\lambda)$, normalized to unity at its peak response. 
The integration is performed over all wavelengths and over 
the area of the detector aperture.
\begin{equation}
 P_\mathrm{src}  =  T_\mathrm{r} \! \! 
\int \! \left[ T_{f}(\lambda)  R_{p}(\lambda) F_{\lambda}(\lambda) \! \!
  \int \! \! \! \! \int \! \! \!
 PSF(x,y,\lambda) \mathrm{d}x \mathrm{d}y \right] \! \mathrm{d}\lambda .
\label{idealequ}
\end{equation}
Wavelength-independent losses due to mirrors are accounted for
by the factor $T_\mathrm{r}$.
The indices $p$ and $f$ indicate dependence on pixel or
filter, respectively.

The filter transmissions (including the out-of-band rejections 
over a wide wavelength range)
and the relative response functions of
the detectors were measured in the laboratory under ``cold"
conditions (at their anticipated operating temperatures in-orbit)
with a Fourier-transform spectrometer (Schubert \cite{schubert93}). 
To calculate $T_\mathrm{r}$, we assumed the
reflectance of each mirror surface to be 
98\,\% and wavelength-independent. Thus we obtain
$T_\mathrm{r}=0.98^n$, with $n$ being the number of mirrors in the
optical paths, which is 6, 7, 7, 6 and 5 for the detectors P1,
P2, P3, C100 and C200, respectively. For P1 an additional multiplicative factor
of 0.95 was included to account for losses due to the
Fabry lens.

We note that the in-band power is only a fraction of the total
incident IR power, since the formula includes the 
wavelength-dependent relative response function of the detector 
material.
Normalizing this response function to 1.0 at its peak
allows us to define the detector responsivity as 
characteristic of the detector pixel, independent of the
filter used. Thus $P_\mathrm{src}$ represents only the fraction of 
the in-band power that actually initiates a photocurrent in the detector 
material.

Since the wavelength dependence of the PSF within a filter band 
is small, we simplified Eq.~\ref{idealequ} to
\begin{equation}
P_\mathrm{src} = A \: f_{\mathrm{PSF}_{f,a}} \: T_\mathrm{r} 
\int  
\left[ T_{f}(\lambda) \: R_{p}(\lambda) \: F_{\lambda}(\lambda)
\right] \mathrm{d}\lambda .
\label{ibflxfnu1}
\end{equation}
by replacing the integral over the aperture and the PSF by 
two factors: (i) The area of the telescope aperture, $A$, and
(ii) the PSF-correction factor, $f_{\mathrm{PSF}_{f,a}}$, dependent on
filter, $f$, and aperture, $a$. This second factor describes -- for
a given combination of filter and aperture -- the fraction of
the PSF that actually enters the detector aperture.

For the initial calculations we used an 
analytical telescope model, taking into account the primary 
and secondary mirrors. Subsequently we replaced
this by a more accurate model (Okumura \cite{okum2000}) based on
numerical Fourier transformation, that also includes effects of 
the support structure of the secondary mirror.
The PSF factors were calculated as the ratio of the in-band 
power resulting from Eq.~\ref{idealequ}, integrated over the 
detector aperture and the total in-band power in the PSF, i.e. 
Eq.~\ref{idealequ} integrated over an infinite aperture.
A $\nu F_{\nu}=const.$ spectrum was adopted as source SED.

Thus far, our intent has been to determine the 
in-band fluxes of known standards to characterise the detectors 
and to calibrate the FCS. 
Once the FCS is calibrated, however, we can use
the instrument to determine the in-band powers of other celestial
sources. To convert these back to flux densities, we must invert
Eq.~\ref{ibflxfnu1}. For regular target observations 
the source SED is generally unknown so we need a common reference
spectrum with a defined colour. We adopt a spectrum of the form
$\nu F_{\nu}=const.$ as Beichman et al. (\cite{beichman88}) did
for IRAS. This enables us to replace
Eq.~\ref{ibflxfnu1} by the much simpler form 
\begin{equation}
P_\mathrm{src} = F_{\nu} \: C1_{f} \: f_{\mathrm{PSF}_{f,a}},
\label{ibflxfnu2}
\end{equation}
where $C1_{f}$ are tabulated factors for the ``reference
wavelength" (see Table~\ref{wavtab}) of each filter. 
The photometric results, $F_{\nu}$,
provided by the off-line processing software (OLP) or the 
interactive analysis (PIA) to the observer are expressed in 
terms of flux densities at the proper reference wavelength of 
the filter used. The choice of the reference wavelength is
arbitrary, as long as the reference spectrum is defined. For
sources with colours deviating from the reference spectrum, a
colour correction must be applied, that can be derived from
Eq.~\ref{ibflxfnu1} and~\ref{ibflxfnu2}. Tabulated colour 
correction factors were derived for a few typical spectra like
blackbodies or power-laws. Table~\ref{wavtab} shows the reference
wavelengths that are used to calculate ISOPHOT flux densities.
\begin{table}
\begin{center}
\begin{tabular}{lccccc}
Filter & $\lambda_\mathrm{ref.}$ & C1 & 
Apert. & lo lim & hi lim \\
	   &        [$\mu$m]  & $[10^{11} \: m^2 Hz]$   &
[$^{\prime\prime}$]  &   [Jy]   &  [Jy]     \\ 
\hline\noalign{\smallskip}
P\_3P29 & 3.3  & 1.473  &  23  & 0.410 &  214.7 \\
P\_3P6  & 3.6  & 7.535  &   "  & 0.156 &  788.7 \\
P\_4P85 & 4.8  & 7.612  &   "  & 0.092 & 1196.9 \\
P\_7P3  & 7.3  & 12.619 &   "  & 0.058 & 1922.8 \\
P\_7P7  & 7.7  & 2.476  &   "  & 0.128 & 4331.8 \\
P\_10   & 10.0 & 4.495  &  52  & 0.041 &  787.4 \\
P\_11P3 & 11.3 & 1.314  &   "  & 0.543 &  371.0 \\
P\_11P5 & 12   & 16.093 &   "  & 0.097 & 1701.6 \\
P\_12P8 & 12.8 & 5.216  &   "  & 0.845 &  389.3 \\
P\_16P0 & 15   & 3.166  &   "  & 0.078 &  706.3 \\
\hline\noalign{\smallskip}
P\_20   & 20   & 5.198  &  79  & 0.696 & 1045.4 \\
P\_25   & 25   & 4.277  &  "   & 1.250 & 1875.9 \\
\hline\noalign{\smallskip}
P\_60   & 60   & 0.591  &  180 & 0.631 &  307.7 \\
P\_100  & 100  & 0.872  &  "   & 2.607 &  719.9 \\
\hline\noalign{\smallskip}
C\_50   & 65   & 0.481  & 43.5 & 0.741 &  276.5 \\
C\_60   & 60   & 0.659  &  "   & 1.036 &   95.8 \\
C\_70   & 80   & 0.697  &  "   & 0.576 &  167.6 \\
C\_90   & 90   & 1.510  &  "   & 0.245 &   47.8 \\
C\_100  & 100  & 0.945  &  "   & 0.094 &   49.5 \\
C\_105  & 105  & 0.611  &  "   & 0.110 &  172.7 \\
\hline\noalign{\smallskip}
C\_120  & 120  & 0.355  & 89.4 & 1.257 &  560.9 \\
C\_135  & 150  & 0.711  &  "   & 1.445 &  363.1 \\
C\_160  & 170  & 1.010  &  "   & 0.837 &  152.6 \\
C\_180  & 180  & 0.537  &  "   & 2.444 &  336.7 \\
C\_200  & 200  & 0.233  &  "   & 1.612 &  532.2 \\
\end{tabular}
\caption{\it The reference wavelengths, conversion factors,
standard apertures and the 
calibrated flux-range for the 25 ISOPHOT filter bands. The
calibrated flux-range is where FCS\,1 was directly calibrated
against external standards. For revolutions prior to 94
sometimes smaller ranges apply. Extrapolations are possible but
may have larger absolute errors. For the C-subsection values
are given for the central pixel of C100 and pixel 1 of C200.}
\label{wavtab}
\end{center}
\end{table}

\subsection{Observations}
The ISO observations to calibrate the internal sources were
scheduled in blocks for each filter band. Each block comprised
measurements of a celestial source, celestial background, heated
FCS and non-heated FCS. The measurements were obtained in
different sequences, but in general the low-flux measurements
were executed before the high-flux measurements to reduce the
time constants of transients. These are shorter for transitions
from low to high flux. The observations for the long-wavelength
detectors P3, C100 and C200 also included direct calibrations of
the redundant FCS\,2. For the other filter bands, FCS\,2 was
calibrated relatively to FCS\,1.

To check the relative effective transmission of filter bands,
measurement blocks of the same detector were further grouped
together, so that responsivity drifts would be minimal during
that time. We verified this by repeating the first measurement 
of the sequence at the end. The results lead to corrections
to the filter transmissions, that are discussed in 
Sect.~\ref{ftof_sec}.

All detector pixels were calibrated individually. Therefore,
measurements of the standard point sources were positioned at
the centre of every pixel. To minimize the measurement time for
the camera arrays, the spacecraft performed a raster mode
pointing sequence with a step size equal to the distance between
pixel centres (46$^{\prime\prime}$ for C100, 92$^{\prime\prime}$
for C200). Larger raster dimensions ($3 \times 5$ for C100 and
$2 \times 7$ for C200) were chosen to obtain the background zero
level also. Where no raster measurement was required,
background measurements were performed in staring mode on 
positions within a 5$^{\prime}$ radius of the source. For
aperture-photometry of faint sources, a one-dimensional raster
of 3 points with the source at the centre position was scanned
back and forth (nodding-mode), to enable elimination of the
baseline drift (see Fig.~\ref{fstab01}).

The use of a raster observation on a fixed sky position was 
also necessary because ISO could neither observe 
positions with a relative offset from a tracked Solar System 
Object (SSO), nor perform raster observations while tracking.
Therefore, all raster observations on SSOs had to be scheduled 
at fixed times with the telescope pointing adjusted to encounter
the moving object.
We made sure that the proper motion of the SSO due to the 
relative motion w.r.t. ISO was always below $60^{\prime\prime}$/hr
during observations. 
The contribution from ISO's motion relative to the centre of the
Earth was only of the order of $\pm 3^{\prime\prime}$/hr,
as science observations were made during the ``slow'' 
part of the orbit.
During the maximum duration of a raster measurement on an
SSO with C100 (688~s), asteroids would travel only
$\approx 12^{\prime\prime}$. 
Taking into account ISO's absolute pointing accuracy for fixed 
targets of about $\pm 2^{\prime\prime}$,
the worst case pointing error amounts to 
$6^{\prime\prime} + 2^{\prime\prime} = 8^{\prime\prime}$, 
which is still much smaller than the large detector apertures or 
pixel sizes used at long wavelengths.

\section{Photometric Corrections} \label{phtcor}
All corrections so far applied to the signal did 
not require quantitative knowledge of the IR flux 
illuminating the detector. However, in this section we will
describe corrections that became necessary after a detailed 
photometric analysis of observations of known celestial 
calibration standards of various brightnesses.

\subsection{Signal Linearisation}
We defined earlier the responsivity, $R$, as a property of the 
detector that is expected to vary in time due to 
cosmic radiation and flux history, but not to depend on
flux level. However, the actual behaviour of 
ISOPHOT detectors was found to be different.

\begin{figure}
\begin{center}
  \vspace{0cm}
  \hspace{0cm}\epsfxsize=88mm \epsfbox{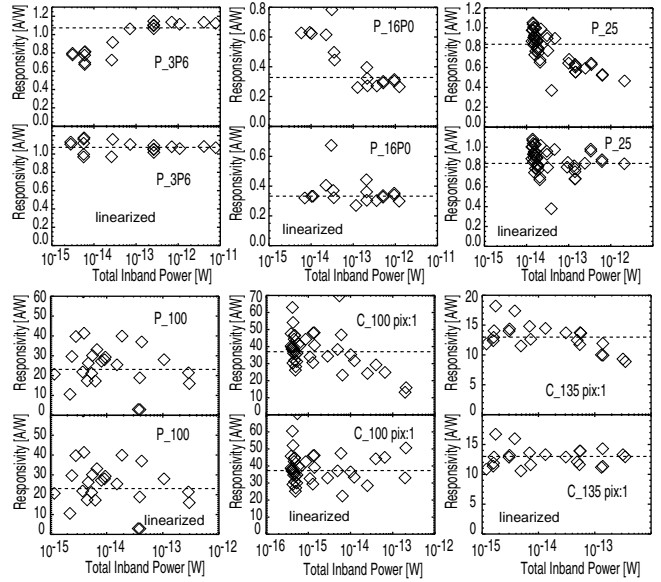}
\caption{\it Naive plots of responsivities against total in-band 
power, assuming linearity (Eq.~\ref{resp1}) for representative
filter bands of all detectors. 
The total in-band power includes the contributions from both  
source and background.
Upper and lower diagrams show the 
responsivities before and after linearisation.
The upper diagrams show a clear 
dependence on in-band power except for P3. 
The remaining scatter in the linearised data is due to the 
variation of the responsivity with time.}
\label{resp_all}
\end{center}
\end{figure}

In its simplest form we calculate the responsivity from the
signals measured on-source $(S_\mathrm{src} + S_\mathrm{bck})$ and off-source
$(S_\mathrm{bck})$ and the PSF-corrected in-band power, $P_\mathrm{src}$,
according to 
\begin{equation}
R = \frac{S_\mathrm{src} \, C_\mathrm{int}}
	{P_\mathrm{src}},
\label{resp1}
\end{equation}
where $C_\mathrm{int}$ is the capacity of the
integrating amplifier. Plotting responsivities from 
measurements of various calibration standards versus in-band
power, however, shows strong correlations 
(see Fig.~\ref{resp_all}). 
Moreover these correlations are different even for filters 
of the same detector subsystem.

If the measured responsivity depends in some way on the
IR flux that falls onto the detector, $R$ can also be
expressed as a function of the photocurrent or the detector
signal. We then rewrite Eq.~\ref{resp1} as
\begin{equation}
R(S_\mathrm{src}+S_\mathrm{bck}) = \frac{(S_\mathrm{src}+S_\mathrm{bck}) \, C_\mathrm{int}}
	{P_\mathrm{src}+P_\mathrm{bck}},
\label{resp2}
\end{equation}
describing the responsivity of the detector when it is pointed
at the position of the source, and further as
\begin{equation}
R(S_\mathrm{src}+S_\mathrm{bck}) = \frac{(S_\mathrm{src}+S_\mathrm{bck}) \, C_\mathrm{int}}
	{P_\mathrm{src}+\frac{S_\mathrm{bck} C_\mathrm{int}}{R(S_\mathrm{bck})}}.
\label{resp3}
\end{equation}
The index $bck$ indicates the signal and in-band power at the
background position. 
In our case, $R(S_\mathrm{bck})$ and $R(S_\mathrm{src}+S_\mathrm{bck})$ are not
necessarily equal and the ratio of the signals
measured on- and off-source is different from the ratio of the
respective in-band powers. To determine $R(S)$ requires 
knowledge about $ R(S_\mathrm{bck})$ because of Eq.~\ref{resp3}.
Considerable errors are introduced 
in comparisons of very different signals if $R$ is assumed to 
be independent of signal.
This applies in particular to 
multi-filter AOTs, multi-aperture AOTs on extended sources, 
maps with large dynamic ranges, and all FCS calibrations where
the signals from FCS and sky are very different. For
measurements on the sky, the biggest systematic errors are
expected when the individual contributions of background and 
source alone are about equal.
\begin{figure*}
\begin{center}
  \vspace{0cm}
  \hspace{0cm}\epsfxsize=180mm \epsfbox{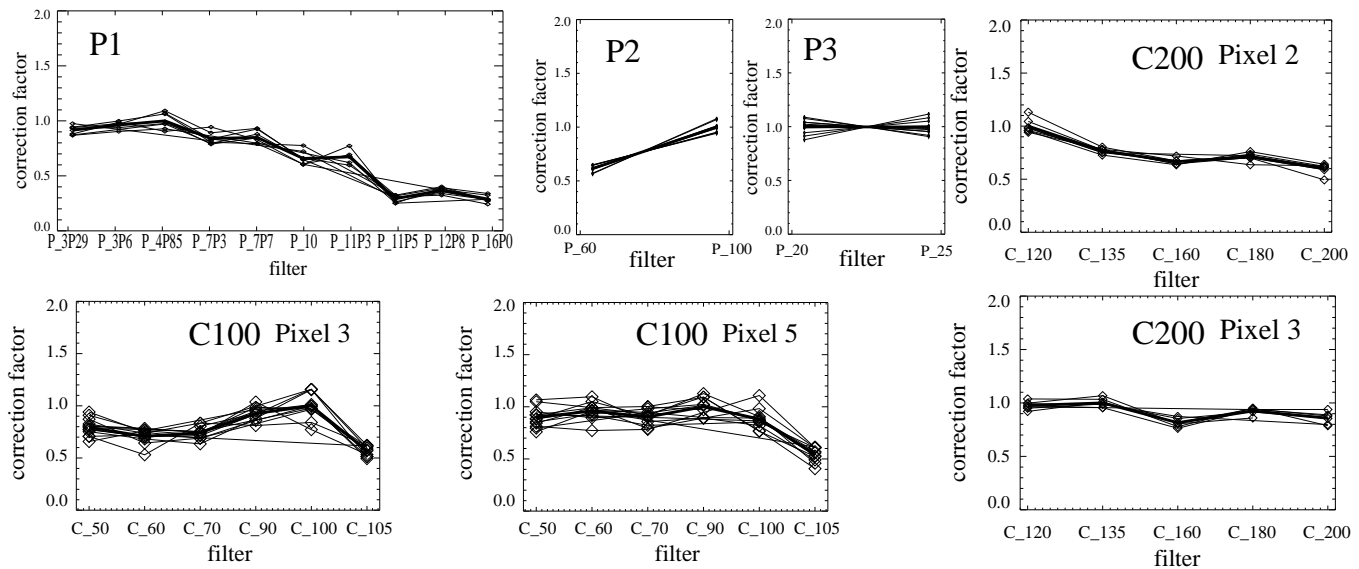}
  \caption{\it Derivation of filter-to-filter corrections.
  We display the data for some representative detectors.
  Each diagram shows the normalized responsivities for the
  indicated detector, derived from various filter sequences. 
  Each sequence was measured on a calibration standard within
  a time interval, sufficiently short to avoid long term drifts.
  Each sequence is connected by a line. The standard deviation
  of the scatter of the relative filter measurements is 2 to
  9\,\% for the P-detectors and 3 to 8\,\% for the C-arrays.}
 \label{ftofdat}
\end{center}
\end{figure*}

All further data reduction steps assume a linear system, so we
decided to linearise the detector signals, $S$, according
to $S^{\prime} = {\cal T}(S)$, where $\cal T$ is a continuously
rising transfer function. The condition for ${\cal T}(S)$ is such that
the results of $S^{\prime}$ inserted into Eq.~\ref{resp1} should
no longer correlate with the detector signal, but should be
distributed around a constant responsivity with minimum
dispersion. The absolute value of this constant is arbitrary
because it cancels out in the flux calibration. We normalized
it to the median of the nonlinearised responsivities calculated
according to Eq.~\ref{resp1}. It should be noted that 
application of the signal linearisation requires the prior 
subtraction of the dark signal.

Although the FCS calibration data comprise a large number of 
measurements, the number of independent responsivities
distributed over the entire flux range per filter band ranges
only between 14 and 28. We developed a manual fitting method to
determine the linearisation tables, by displaying the
results of Eq.~\ref{resp3} and interactively modifying
the linearisation function, ${\cal T}(S)$.
This approach led to a
better exploration of parameter space, a smoother 
transition at the boundaries, and a more tolerant treatment of
outliers without the need to develop sophisticated fitting
algorithms. We increased the objectivity of the method
by keeping the shape of $\cal T$ simple, generally 
close to a polynomial of second or third order. We also minimised
differences between pixels of the array detectors by fitting
all pixels of one detector in parallel on the same
screen. In spite of the large scatter of responsivities,
simple functions $\cal T$ were found that significantly
increase the consistency of data measured at different flux
levels.

Signal linearisation is performed by interpolation in
a lookup table spanning the full range of possible signals.
This table is calibrated only in the range covered by
calibration standards but, in practice, signals occur
outside this range that are higher, lower, or even negative in
the case of noisy measurements close to the dark current.
Reasonable entries for these cases had to be defined. For
high signals beyond the point where data are available,
we maintained a constant responsivity.
Small signals are constrained so that 
an already dark-subtracted zero signal is not 
changed by signal linearisation. Therefore, we continue 
${\cal T}(S)$ between the smallest point in
the lookup table that is still determined by valid data, and
zero, by a straight line. This assumes that the 
responsivity does not vary in that range.
c) The most difficult cases are negative signals, because
negative fluxes are simply undefined and appear only as a
result of noise. The number of such cases actually encountered
is low due to the reduced noise after signal averaging
over periods of constant flux (SCP level). We decided to continue
the transfer function such that ${\cal T}(S) = -{\cal T}(-S)$,
to cover all situations that occur.

\subsection{Filter-to-Filter Calibration}
\label{ftof_sec}
Calculating the detector responsivity from standard source 
observations according to
\begin{equation}
R^{\prime}_{p,f}=\frac{(S_{\mathrm{src}_{p,f}} - S_{\mathrm{bck}_{p,f}}) \, 
C_\mathrm{int}}{P_{f}}, 
\label{resp_1}
\end{equation}
after application of all signal corrections described so far, 
still yields different results for the same detector pixel,
$p$, measured in different filters, $f$, during the same batch
of observations. 
The ratios of responsivities between filters are reproducible 
for different calibration sources and cannot 
be explained by responsivity variations with time. 

The worst case is detector P1, where $R$ varies by a factor
of 4 between filters P\_3.29 and P\_16 (see Fig.~\ref{ftofdat}). 
Currently, the wavelength-dependent flux losses of this detector
remain unexplained.
We speculate, however, that the sapphire substrate that 
mechanically supports the detector crystal within the 
kidney-shaped aluminium cavity might play a role.
Sapphire is transparent in the NIR and starts to absorb beyond
about 6~$\mu$m onwards, in the middle of P1's
wavelength range (Tropf \& Thomas \cite{tropfthom98}). The absorption peak
is at $\approx$17.5~$\mu$m. Some fundamental lattice 
vibration modes occur in the reststrahlen region 
above 13~$\mu$m. 
Top and bottom of both detector and substrate are gold-coated. 
The remaining effective area of the detector, where 
radiation from within the cavity can enter, is only twice that 
of the substrate. Given the high refractive index of Si, around 
3.4, and the consequently higher reflectivity compared
to Al$_{2}$O$_{3}$ (refractive index of only 1.7), we estimate that 
about the same number of photons enters both detector and 
substrate.
The cavity was designed to counteract the high refractive 
index of the detector material by maximizing the number of 
reflections within it. Assuming that the sapphire substrate 
turns ``dark" at longer wavelengths, it is conceivable that,
with every reflection, a substantial fraction of the radiation
enters the substrate and is attenuated. The effective response 
of the detector system would thus deteriorate. 
Another IR absorber present within the cavity is glue (Stycast).
Unfortunately the question of P1's quantum efficiency remains
largely academic, so that the effort of a much more detailed 
analysis is not justified.

The other detectors show less dramatic variations by factors up
to 2. Moreover, the ratios found between different filters are 
not the same for different pixels of the same C-detector array. 
We attribute this to a projection of spatial nonuniformities 
of the filter surface onto the detector array,
because the ISOPHOT filters are not located at the pupil of the 
optical path, but close to the detectors.
The overall variation between filters of all detectors, except 
P1, is most likely due to 
optical misalignments and diffraction effects that are 
neglected in our ideal instrument model.

We applied a correction to the responsivity by introducing a 
matrix, $\chi_{p,f}$. It works equally for P- and C-detectors.
To obtain a wavelength independent detector responsivity, we 
split $R^{\prime}$ of Eq.~\ref{resp_1} 
into filter-independent and filter-dependent parts, 
so that $R_{p} \, \chi_{p,f} = R^{\prime}_{p,f}$.
The matrix, $\chi_{p,f}$, was established from multi-filter 
measurements 
on calibration standards, executed so that the 
change of responsivity with time could be neglected for 
a filter sequence.
After eliminating poor quality measurements, the responsivities 
measured for a detector pixel, $p$, in a filter sequence were 
renormalized. The factors per sequence were chosen to 
achieve the best match among all sequences for a given pixel.
After calculating the average responsivities, $R_{p,f}$, per filter, $f$,
we normalized to the maximum responsivity found for that 
pixel using $\chi_{p,f} = R_{p,f} / \mathrm{max}(R_{p,f})$, 
thus assuming that the matrix 
describes flux losses w.r.t. the ideal model.
The normalized data and the resulting filter-to-filter 
corrections, $\chi_{p,f}$, for some detector pixels and filter bands,
are shown in Fig.~\ref{ftofdat}.
The standard deviation of the scatter of the relative
filter measurements is 3 to 8\,\% for the C-arrays and
2 to 9\,\% for the P-detectors.

\section{Derivation of FCS power curves} \label{fcspow}
\begin{figure}
\begin{center}
 \vspace{0cm}
  \hspace{0cm}\epsfxsize=88mm \epsfbox{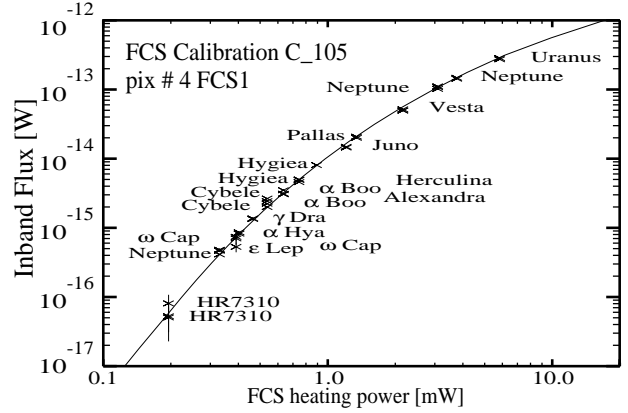}
  \caption{\it Constructing a power curve from calibration data. 
  The names indicate the sources that have been used to calibrate 
  the respective data points.
  }
 \label{fcscrv0}
\end{center}
\end{figure}
The FCSs were calibrated against celestial standards
in all 25 filter bands. These measurements yield the relationship
between in-band power, $P$, and FCS heating power, $h$, for all
combinations of filter band, $f$, detector pixel, $p$, and FCS.
Signals are measured differentially. The source signal is obtained
from the difference between on-source, $S_\mathrm{src}$, and off-source,
$S_\mathrm{bck}$, signals while the FCS signal is calculated as the
difference between the signals measured with heated FCS($S_\mathrm{FCS}$)
and cold FCS($S_\mathrm{str}$). 
Since the measurements of FCS and celestial standard
were performed in the same filter, the responsivity does not 
need to be explicitly calculated. 
We used
\begin{equation}
P_\mathrm{FCS}^{\prime}(h) = \frac{(S_\mathrm{FCS}(h)-S_\mathrm{str})\: 
                P_\mathrm{src}}{S_\mathrm{src}-S_\mathrm{bck}},
\label{sigratios}
\end{equation}
where $P_\mathrm{FCS}^{\prime}(h)$ is the in-band power incident on the
detector from the FCS and $P_\mathrm{src}$ is defined in 
Eq.~\ref{ibflxfnu1}. Fig.~\ref{fcscrv0} shows an example
of the power curve
resulting from observations of different calibrators for 
pixel~4 in the C\_105 filter band.

\subsection{Illumination Matrices and FCS Apertures}
\begin{figure}
\begin{center}
  \vspace{0cm}
  \hspace{0cm}\epsfxsize=88mm \epsfbox{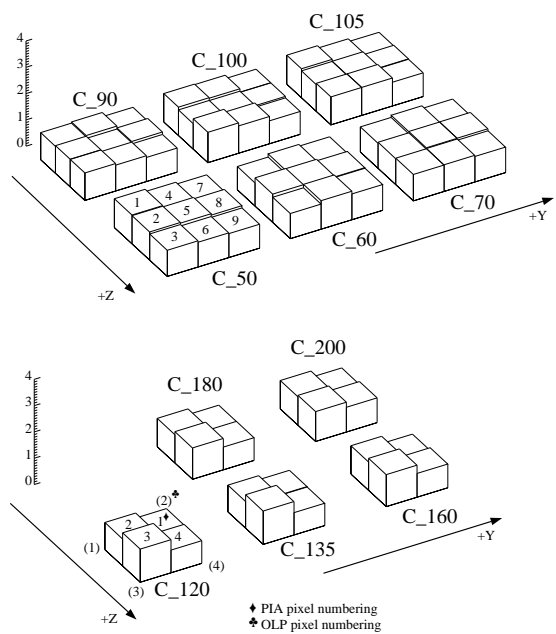}
  \caption{\it Relative illumination of the array detectors C100
  and C200 (illumination matrices) by FCS\,1 (valid after rev. 93). 
  Pixel numbers and ISO coordinate axes are given.}
  \label{illumat}
\end{center}
\end{figure}
\begin{figure*}
\begin{center}
  \vspace{0cm}
  \hspace{0cm}\epsfxsize=88mm \epsfbox{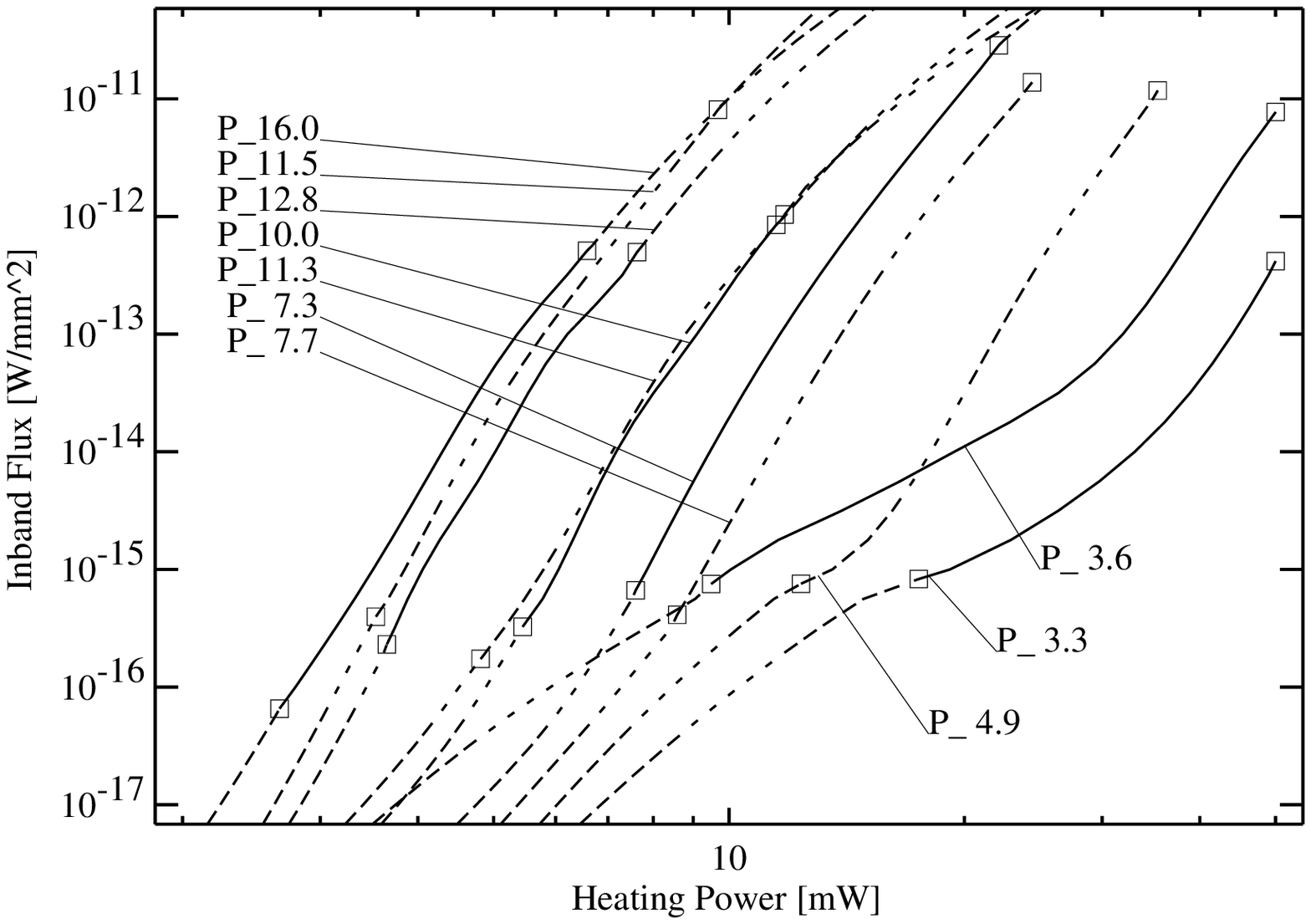}
  \hspace{0cm}\epsfxsize=88mm \epsfbox{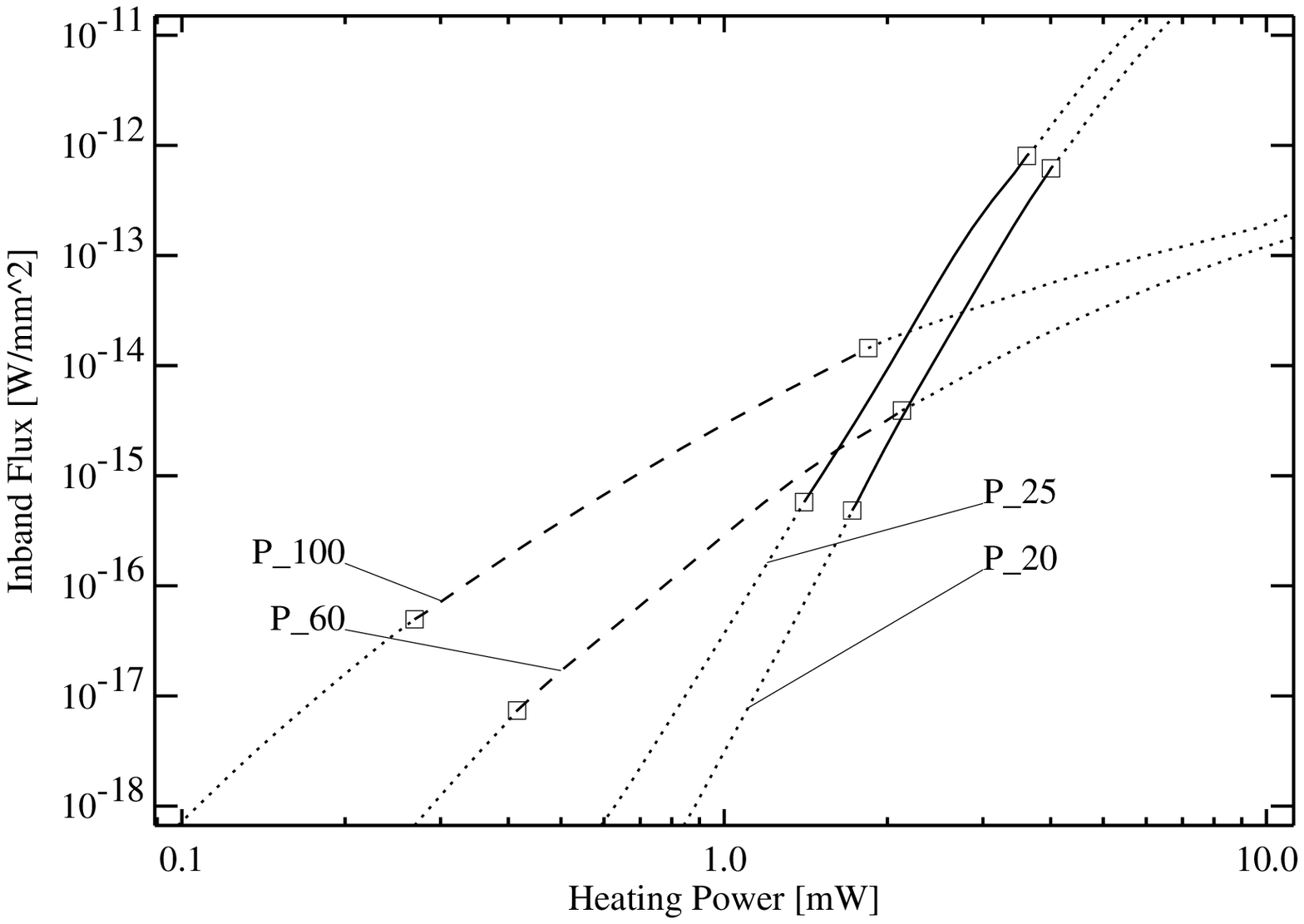}
  \hspace{0cm}\epsfxsize=88mm \epsfbox{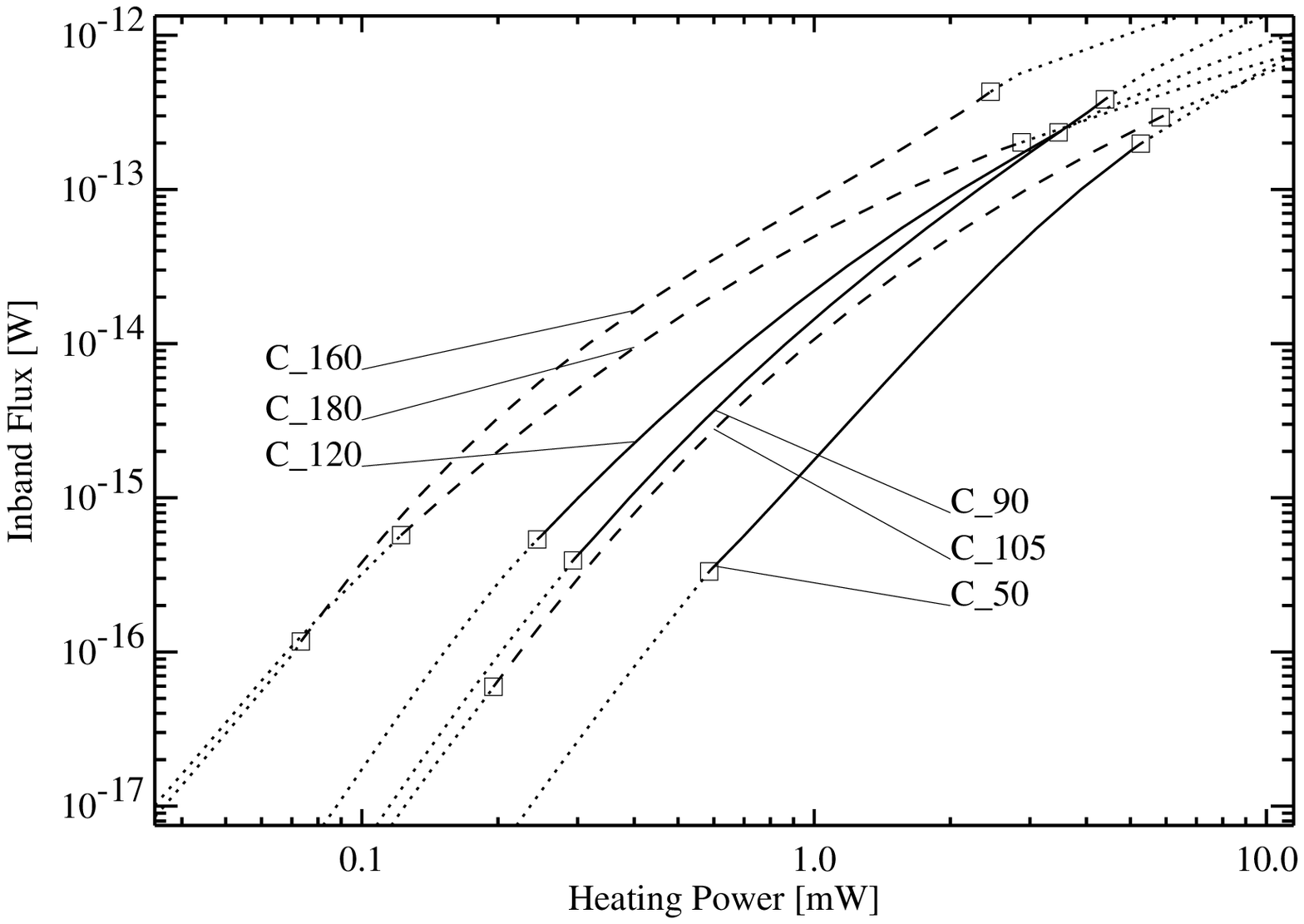}
  \hspace{0cm}\epsfxsize=88mm \epsfbox{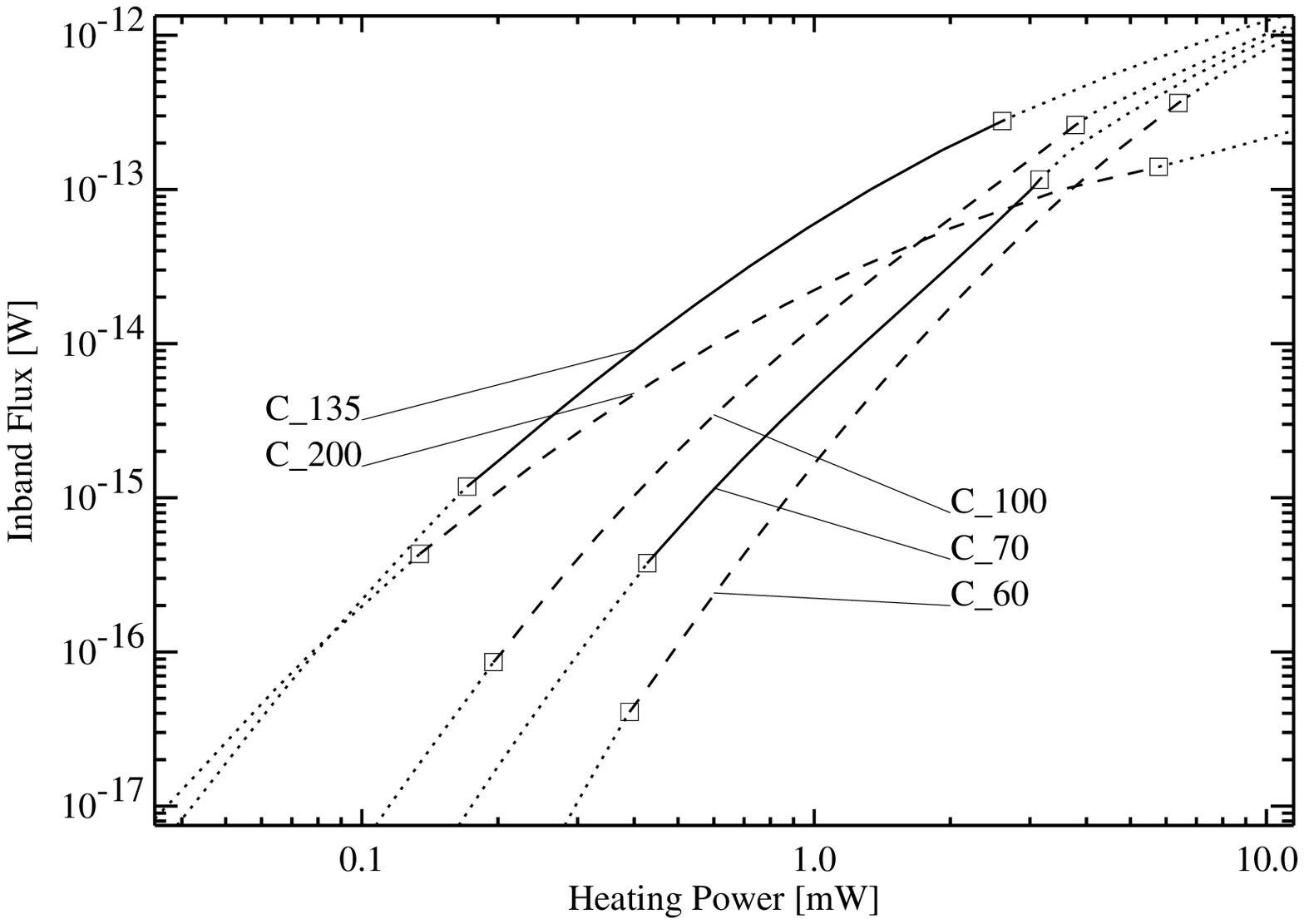}
  \caption{\it FCS power curves of all filter bands. 
  For P3, C100 and C200 we show the versions valid 
  after the change of FCS\,1/TRS\,2 in rev. 94. 
  The interpolated parts are indicated by solid or dashed lines
  with small squares at the boundaries,
  while the model extrapolations are indicated by 
  dotted lines.
   }
  \label{fcscrv1}
\end{center}
\end{figure*}
\begin{figure*}
\begin{center}
  \vspace{0cm}
  \hspace{0cm}\epsfxsize=5.8cm \epsfbox{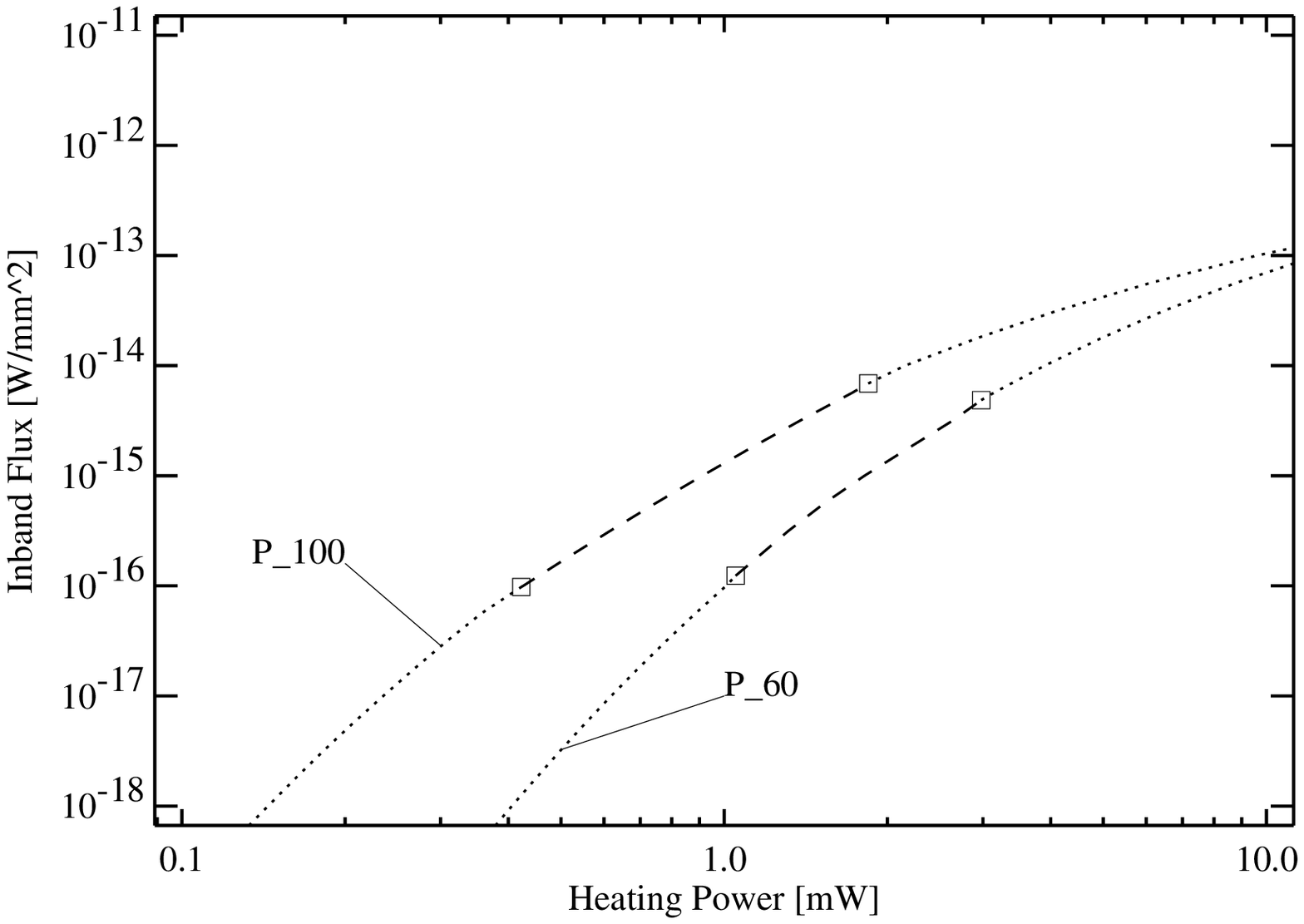}
  \hspace{0cm}\epsfxsize=5.8cm \epsfbox{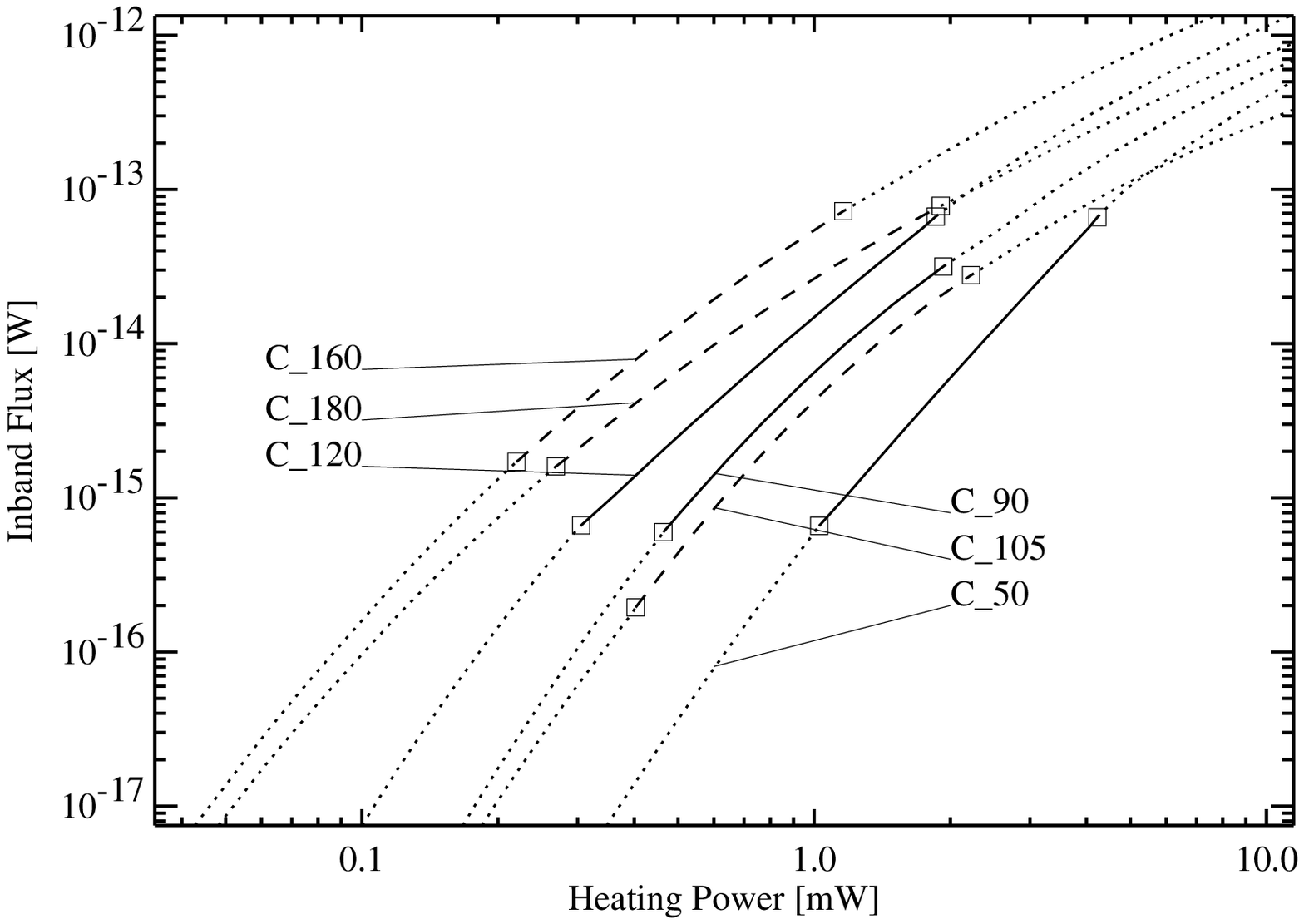}
  \hspace{0cm}\epsfxsize=5.8cm \epsfbox{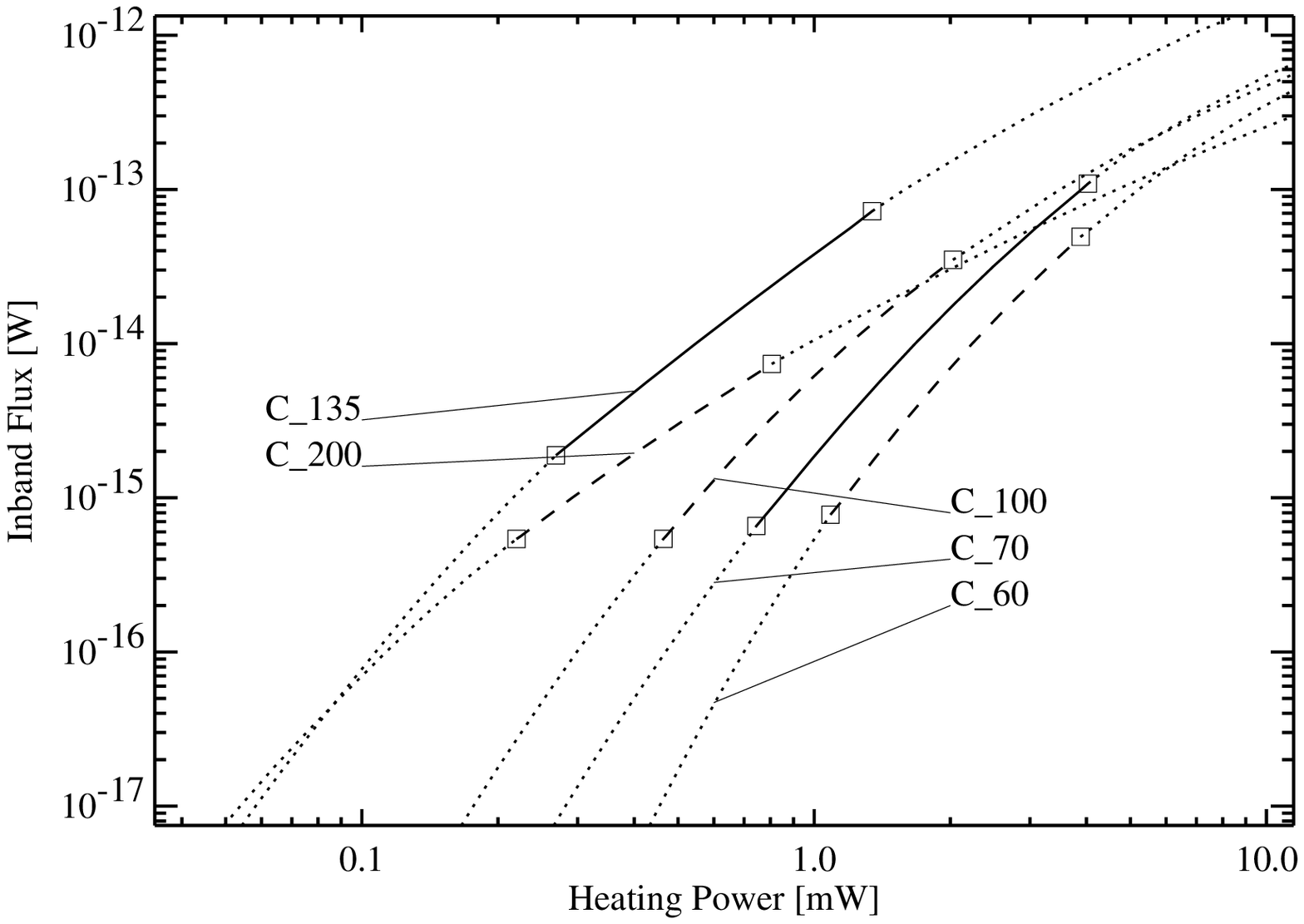}
  \caption{\it FCS power curves of all filter bands of the 
  long wavelength detectors before the change in 
  FCS\,1/TRS\,2.}
  \label{fcscrv2}
\end{center}
\end{figure*}
FCS power curves for different pixels of the same C-array differ
by constant factors. 
These factors result from both the nonuniform illumination of
the detector arrays by the FCS and the putative spatial inhomogeneities
in the filter. For each filter, these factors constitute the 
illumination matrix, $\Gamma_{p,f}$, as shown in Fig.~\ref{illumat}. 
Thus the FCS curve for each filter is split into
a pixel-independent part, $P_{\mathrm{FCS}_{f}}(h)$, 
and a pixel-dependent factor, $\Gamma$,
so that $P^{\prime}_{\mathrm{FCS}_{p,f}}(h) = \Gamma_{p,f} \, P_{\mathrm{FCS}_{f}}(h)$. 
The FCS curve is normalized such that $\Gamma_{p,f}$, averaged over 
all pixels becomes 1, for a given filter.
The resulting FCS power curves for FCS\,1 are shown in 
Figs~\ref{fcscrv1} and \ref{fcscrv2}. 
Between revolution 93 and 94 the TRS\,2 of FCS\,1, which was
routinely used with the long-wavelength detectors, abruptly increased 
its brightness by about a factor of 2. This change necessitated a
recalibration of the corresponding filter bands and
resulted in a second set of FCS calibration tables for 
P3, C100 and C200.
Fig.~\ref{fcscrv2} shows the FCS power curves that apply to the 
period before revolution 94.
The non-uniformity of the FCS illumination also affects the
FCS signal measured in different apertures of the P-detectors.
For a fixed heating power the FCS signal is not proportional to
the area of the aperture. Strong deviations exist for apertures 
much larger or smaller than the standard apertures. The
treatment of these cases is still poorly understood and is
beyond the scope of this paper.
For use with standard and very similar apertures,
the in-band powers of the PHT-P power curves were divided by 
the area of the standard aperture of the 
measurement, and are expressed in units of W/mm$^2$.

\subsection{Extrapolation}
During the mission, FCS heating powers were used that fell
beyond the calibrated range of heating powers. We used 
the grey body model described in Sect.~\ref{calstrat}
to extrapolate the power curves.
Including the spectral transmission of the beam splitter of
the FCS and replacing the celestial source SED by the grey body spectrum,
the in-band powers are calculated in accordance with Eq.~\ref{ibflxfnu1}.
The free parameters $\alpha$ and
$\beta$ and the effective solid angles of the detector subsystems
are obtained from
least-squares fits to the FCS curves (Schulz \cite{schulz93}). 
Comparison of the model to the in-orbit data shows good 
consistency, especially at longer wavelengths. At shorter 
wavelengths the consistency is poorer.
Extending the measured FCS power curves by the models
provided the values outside the range
of measured heating powers. This range is flagged in the
associated calibration tables.

\subsection{FCS Straylight}
\label{fcsstray}
Ground calibrations already showed an increased signal compared
to the dark signal when measuring the signal from a ``cold" FCS.
Considering the internal ISOPHOT geometry, this additional signal
is most likely picked up from radiation within the field of view
of the telescope (Schulz \cite{schulz93}).
The in-orbit observations, where the cold FCS signal was found to 
be correlated with the background signal, confirm the presence of
this kind of straylight for C200 (Cornwall \cite{cornw97}). 
C100 and P3 show no detectable straylight component. P2 shows a
constant value of 0.02-0.04~V/s for a signal $>$ 0.2~V/s (on source)
and a steep fall-off for lower signals. 
Straylight in P1 is always below 0.07~V/s but
shows a nonlinear dependence on signal.
The nonlinear dependences for both detectors P1 and P2 are quite 
uncertain but appear to be due to the varying dark signal 
and systematic errors in the 
transient fit at low signals, 
rather than to straylight, which is expected to be proportional 
to the intensity of the incident flux components.
We conclude, therefore, that FCS straylight should be taken into 
account for C200, but can be neglected for the other cases. We 
determined the straylight at the FCS position measured with C200 
to be 5\,\% of the background level.

\section{Flux Calibration} \label{flxcal}
We can now detail the procedure to
calibrate ISOPHOT observations.
First, the detector responsivity is determined from the 
FCS measurement using
\begin{equation}
R_{p} = \frac{(S_{\mathrm{FCS}_{p,f}}- S_{\mathrm{str}_{p,f}})C_\mathrm{int}}
             {P_{\mathrm{FCS}_{f}}(h) \Gamma_{p,f} \chi_{p,f}}.
\label{resp_fcs}
\end{equation}
$P_{\mathrm{FCS}_{f}}(h)$, $\Gamma_{p,f}$ and $\chi_{p,f}$ are taken from 
the calibration tables. The FCS straylight signal $S_{\mathrm{str}_{p,f}}$
was actually measured only in absolute photometry AOTs and 
in dedicated FCS calibration measurements. In all other cases
it is replaced by the time-dependent dark signal 
(see Sects.~\ref{darksect} and \ref{fcsstray}).
C200 should be corrected for the 5\,\% straylight contribution of 
the background.
Second, the in-band power, $P_{f}$, for all filters of the 
detector subsystem is calculated as
\begin{equation}
P_{p,f} = \frac {S_{\mathrm{src}_{p,f}}  C_\mathrm{int}}
                 {R_{p} \chi_{p,f}}.
\label{flux_src}
\end{equation}
The transformation to a flux density, $F_{\nu}$, at the 
reference wavelength of the filter (e.g. 12~$\mu$m for the 
P\_11.5 filter) is done by changing
Eq.~\ref{ibflxfnu2} to
\begin{equation}
F_{\nu_{p,f}} = \frac{P_{p,f}}{ C1_{f} \: f_{\mathrm{PSF}_{f,a}}}.
\label{ibflxfnu3}
\end{equation}
The flux density of the point source is obtained by subtracting
the flux density derived from a 
measurement on the background, following the same 
calibration steps as for the source observation. 
The resulting flux density is calculated at the reference 
wavelength of the filter for an assumed 
$\nu F_{\nu} = const.$ spectrum and still has to be 
colour-corrected in accord with the actual source spectrum.
It should be noted that, when the point source is not placed at 
the centre of the detector pixel or aperture, a different 
PSF correction factor must be applied.
In particular, in observing modes with C-detectors, the sum of 
the fluxes found in all pixels can be calculated and a 
PSF correction factor, $f_{\mathrm{PSF}_{f,a}}$,
determined for the entire array, is applied.

\subsection{Accuracy}
\begin{table}
\begin{center}
\begin{tabular}{lccccc} 
Filter & n & $max$ & $\sigma$&$\sigma_\mathrm{lo}$&$\sigma_\mathrm{hi}$\\
 & & [\%] & [\%]	 &	[\%]	 & [\%] \\
\hline\noalign{\smallskip}
 P\_3P29 		 &  5  &   6 &  4 &  4  &  - \\
 P\_3P6  		 & 13  &  17 &  6 &  8  &  1 \\
 P\_4P85 		 & 10  &  27 & 10 &  -  &  3 \\
 P\_7P3  		 & 19  &  26 &  8 & 11  &  4 \\
 P\_7P7  		 & 17  &  15 &  5 &  5  &  5 \\
 P\_10   		 & 11  &  21 &  8 &  -  &  6 \\
 P\_11P3 		 &  6  &  15 &  8 &  -  &  8 \\
 P\_11P5 		 & 26  &  42 & 13 & 18  &  4 \\
 P\_12P8 		 & 15  &  19 &  8 &  9  &  7 \\
 P\_16P0 		 & 16  &  40 & 15 & 12  & 17 \\
\hline\noalign{\smallskip}
 P\_20  		 & 17  &  14 &  7 &  4  &  4 \\
 P\_25  		 & 32  &  19 & 10 &  9  & 10 \\
\hline\noalign{\smallskip}
 P\_60  		 & 75  &  82 & 20 & 22  & 12 \\
 P\_100 		 & 27  &  52 & 25 & 27  & 22 \\
\hline\noalign{\smallskip}
 C\_50  		 & 372 &  53 & 15 & 14  & 16 \\
 C\_60  		 & 330 &  48 & 13 & 15  & 12 \\
 C\_70  		 & 332 &  44 & 15 & 11  & 16 \\
 C\_90  		 & 250 &  76 & 18 & 15  & 20 \\
 C\_100 		 & 621 &  94 & 18 & 19  & 14 \\
 C\_105 		 & 295 &  76 & 11 & 14  &  7 \\
\hline\noalign{\smallskip} 
 C\_120 		 & 154 &  22 &  9 & 10  &  7 \\
 C\_135 		 & 158 &  40 &  9 & 10  &  9 \\
 C\_160 		 & 239 &  57 & 16 & 18  & 11 \\
 C\_180 		 & 157 &  31 &  8 &  8  &  8 \\
 C\_200 		 & 153 &  45 &  8 & 10  &  6
\end{tabular}				 
\caption{\it Comparison of derived FCS power curves with measured 
calibration data. The deviations are given in percent. n is the
number of measurements used, $max$ is the maximum deviation
found and $\sigma$ is the r.m.s. deviation. The last two
columns $\sigma_\mathrm{lo}$ and $\sigma_\mathrm{hi}$ give the r.m.s.
deviations in the low and high flux regions separated by the
vertical dashed lines in Fig.~\ref{cmpfcsP} and \ref{cmpfcsC}.\label{tabaccrt}}
\end{center}

\end{table}
\begin{table}
\begin{center}
\begin{tabular}{lccccc} 
Filter & n & $max$ & $\sigma$ &$\sigma_\mathrm{lo}$&$\sigma_\mathrm{hi}$\\
 & & [\%] & [\%] & [\%] & [\%]	\\
\hline\noalign{\smallskip}
 P\_60  		 &  13 & 54 & 21 & 16 & 22 \\
 P\_100 		 &   8 & 10 &  7 &  - &  8 \\
\hline\noalign{\smallskip}  
 C\_50  		 & 106 & 35 & 18 & 15 & 19 \\
 C\_60  		 &  92 & 58 & 22 & 25 & 19 \\
 C\_70  		 & 108 & 29 & 14 & 13 & 14 \\
 C\_90  		 & 109 & 31 & 17 & 20 & 14 \\
 C\_100 		 &  94 & 39 & 16 & 18 & 12 \\
 C\_105 		 &  92 & 41 & 17 & 16 & 17 \\
\hline\noalign{\smallskip} 
 C\_120 		 &  49 & 17 &  7 &  7 &  6 \\
 C\_135 		 &  41 & 16 &  6 &  6 &  7 \\
 C\_160 		 &  58 & 23 &  8 &  8 &  8 \\
 C\_180 		 &  35 & 13 &  5 &  4 &  6 \\
 C\_200 		 &  27 & 15 &  7 &  7 &  -
				      
\end{tabular}		 
\caption{\it Same as previous table but for the period before 
rev. 94.\label{tabaccpv}}
\end{center}

\end{table}
%
\begin{figure*}
\begin{center}
 \vspace{0cm}
 \hspace{0cm}\epsfxsize=88mm \epsfbox{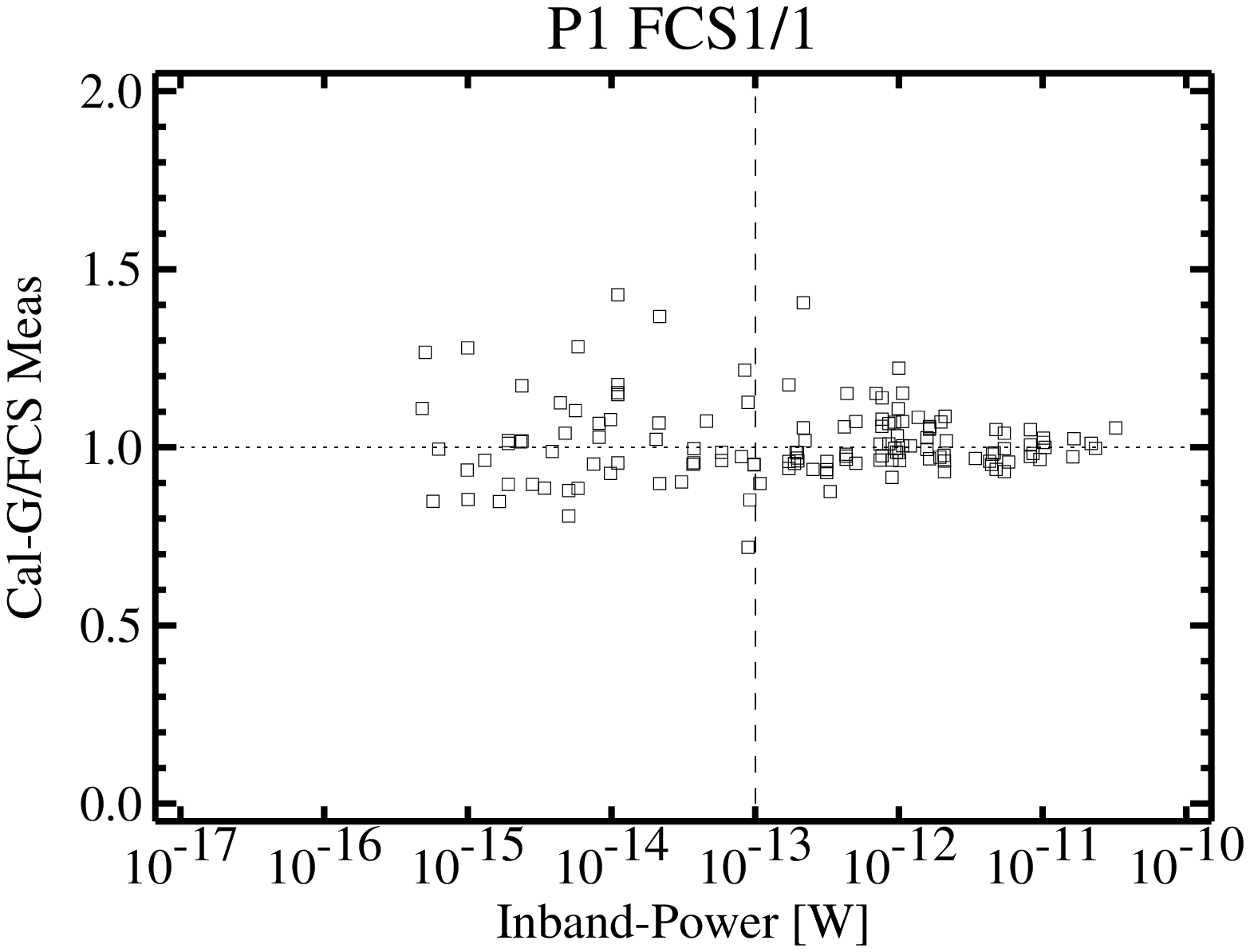}
 \hspace{0cm}\epsfxsize=88mm \epsfbox{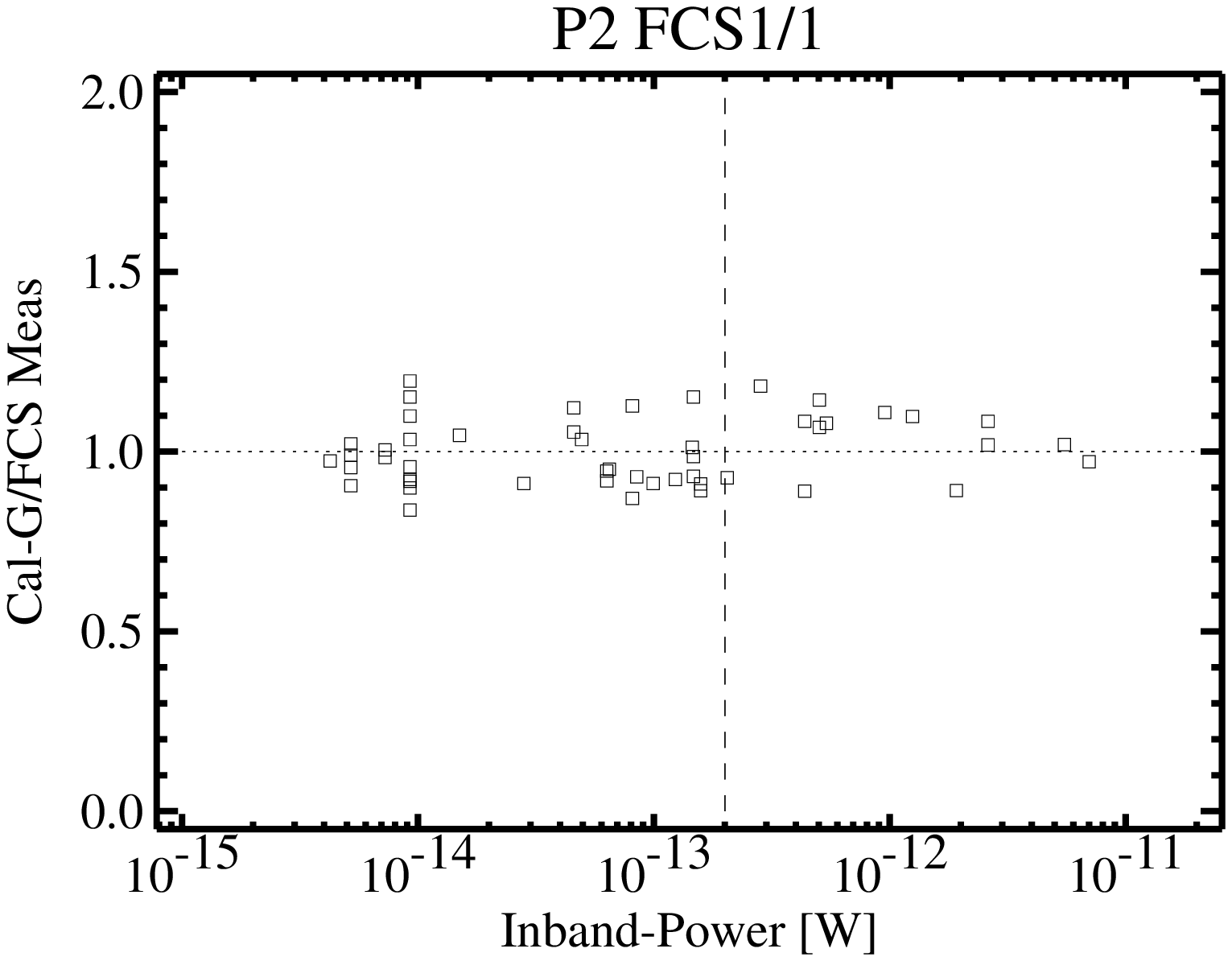}
 \hspace{0cm}\epsfxsize=88mm \epsfbox{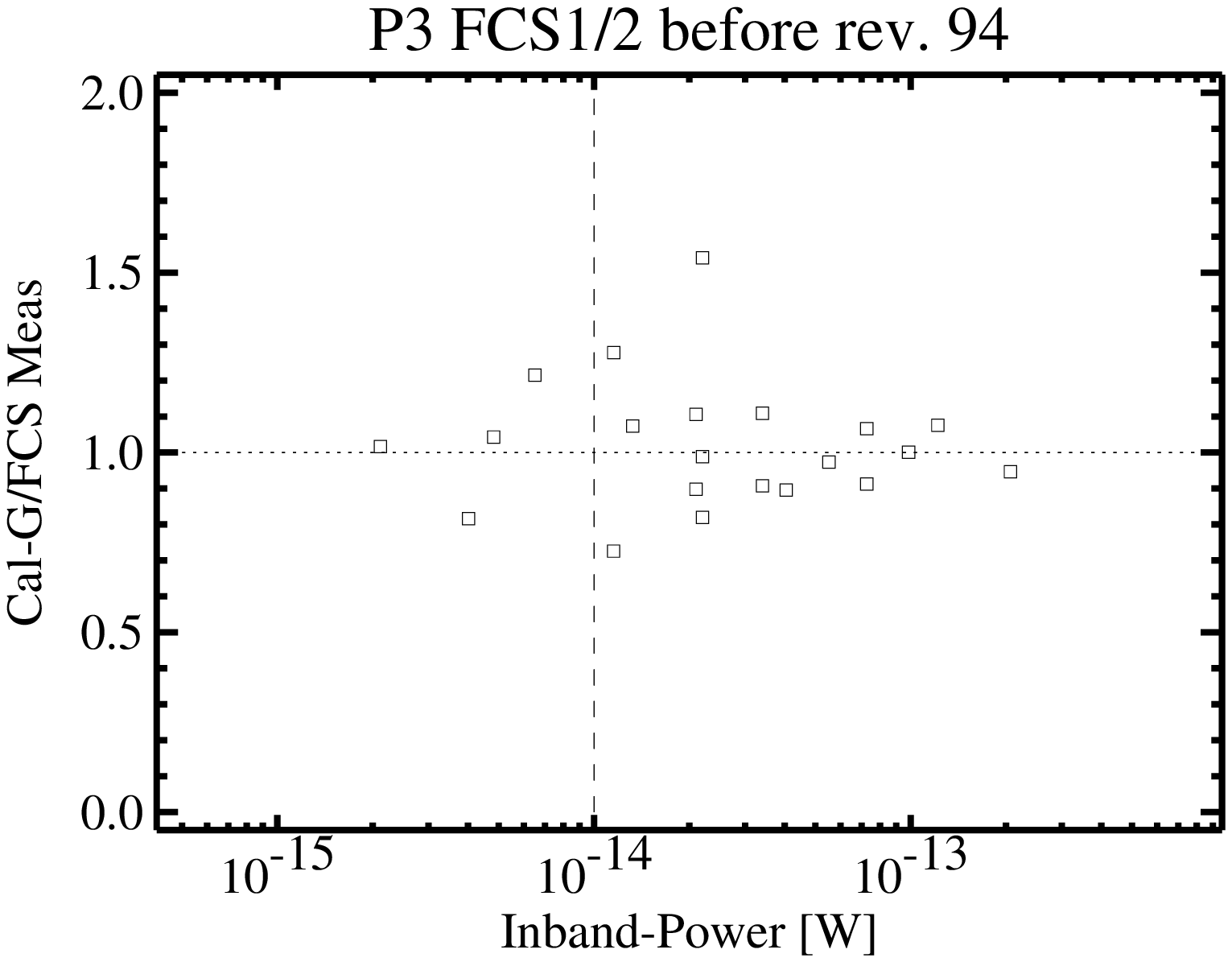}
 \hspace{0cm}\epsfxsize=88mm \epsfbox{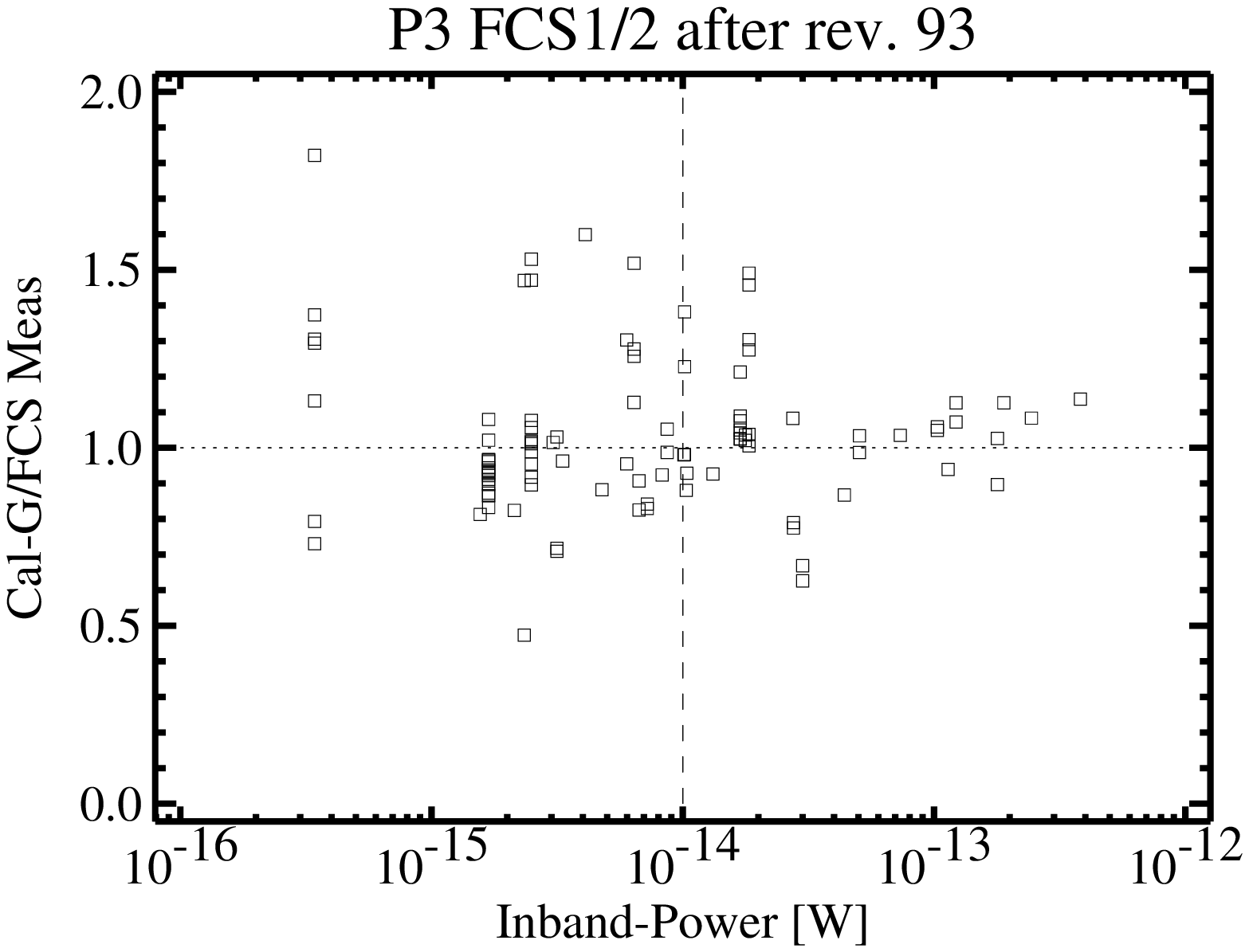}
 \caption{\it Calibration accuracy derived from calibration 
 consistency. The diagrams show the ratio of the 
 FCS power curves times the illumination matrices and the 
 data points derived for the
 P-detectors, plotted against FCS in-band power. 
 The residuals of all filters
 of a subsystem are combined within one plot. The scatter is a
 measure of the error budget for single observations. The
 vertical dashed line separates the low and high flux intervals.
 }
 \label{cmpfcsP}
\end{center}
\end{figure*}
%
\begin{figure*}
\begin{center}
 \vspace{0cm}
 \hspace{0cm}\epsfxsize=88mm \epsfbox{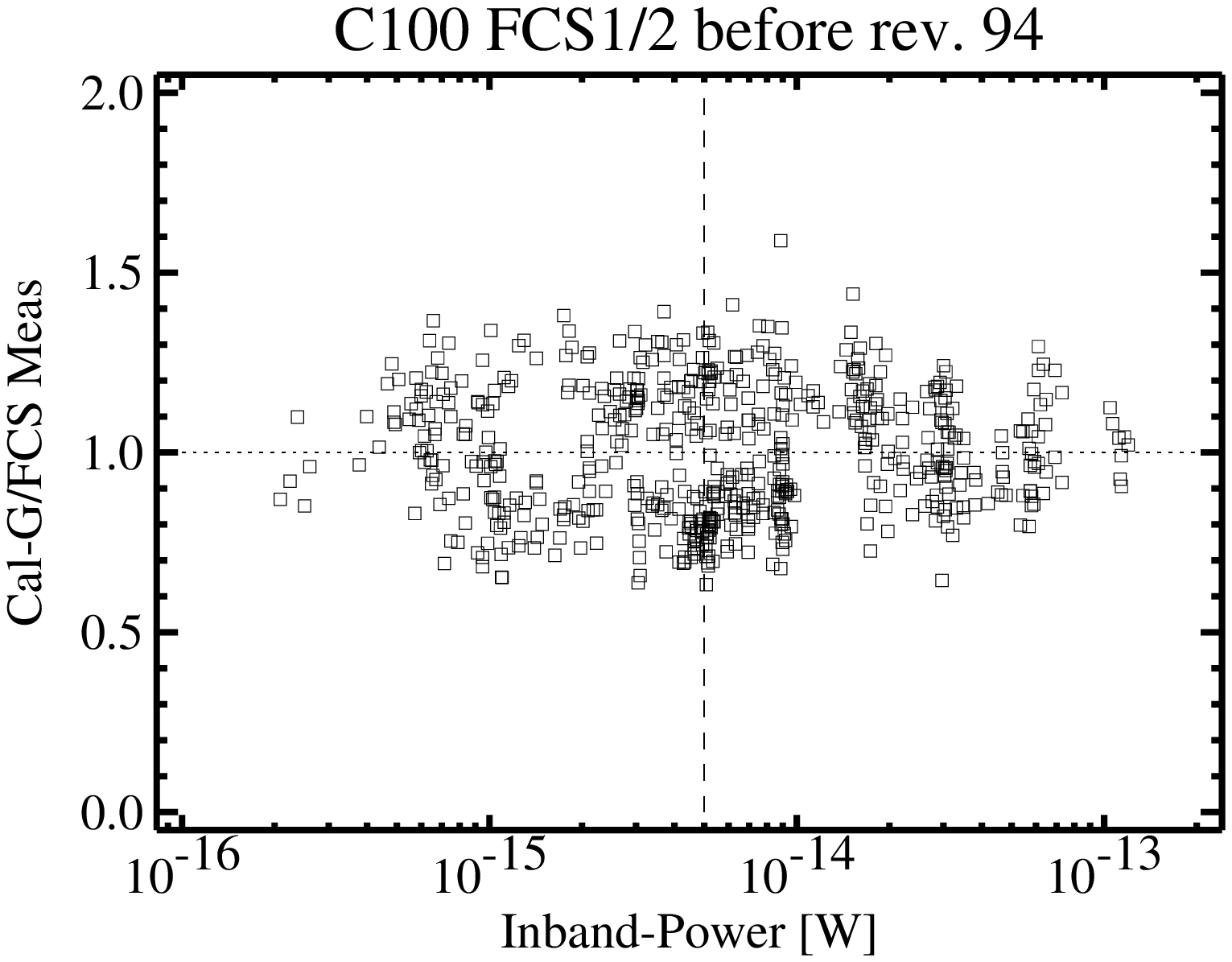}
 \hspace{0cm}\epsfxsize=88mm \epsfbox{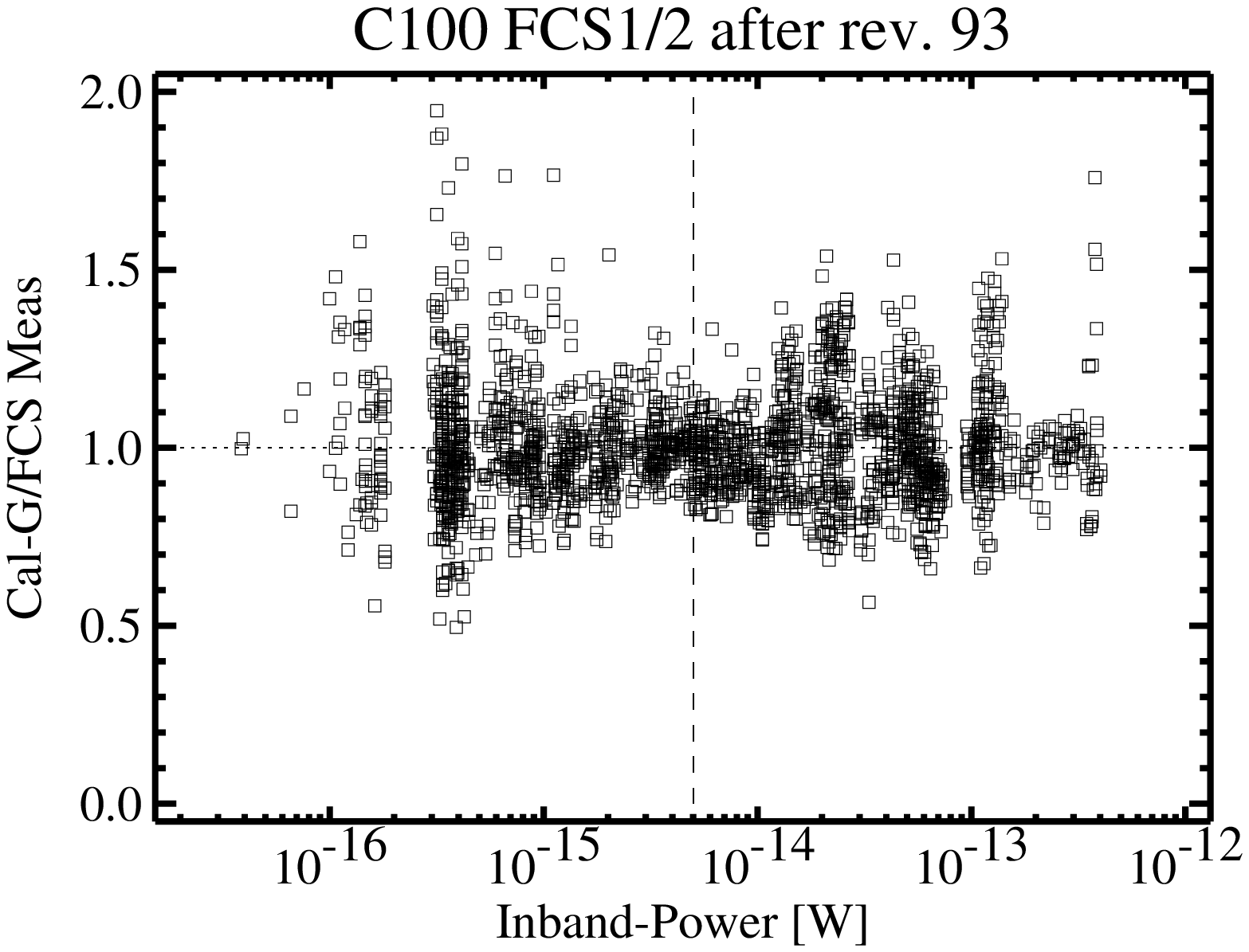}
 \hspace{0cm}\epsfxsize=88mm \epsfbox{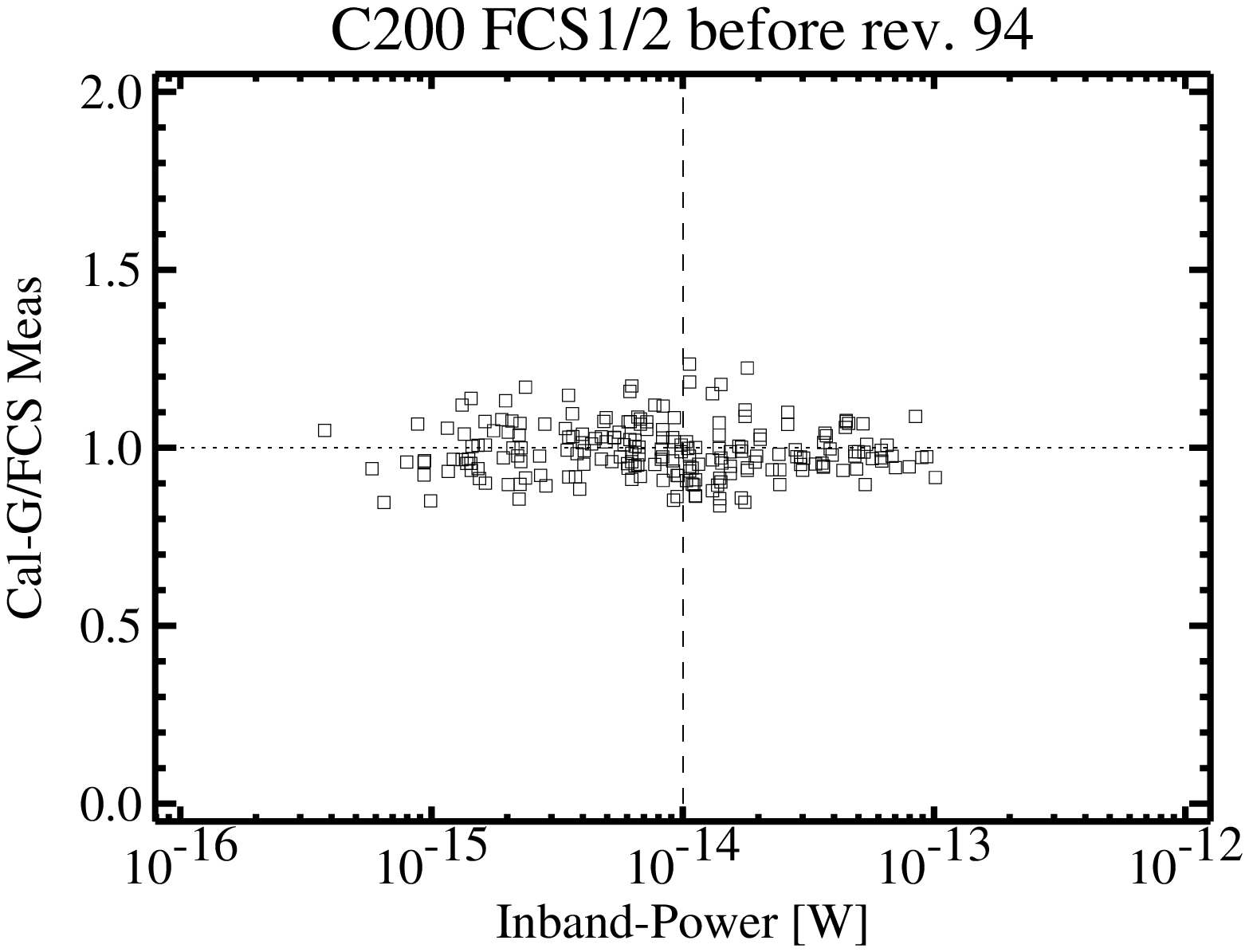}
 \hspace{0cm}\epsfxsize=88mm \epsfbox{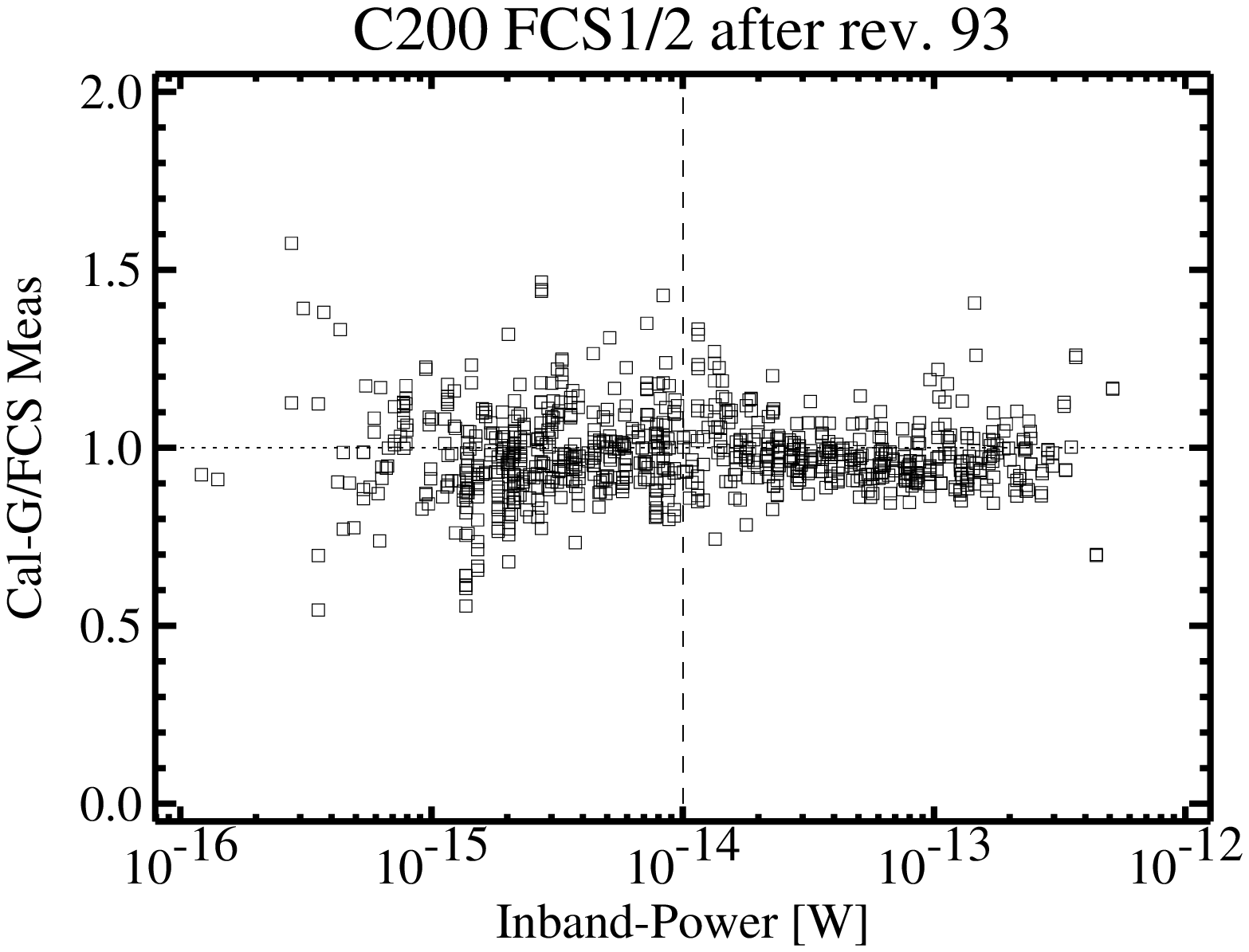}
 \caption{\it Calibration accuracy derived from calibration 
 consistency. The diagrams show the ratio of the 
 FCS power curves and the data points derived for
 the C-detectors, plotted against FCS in-band power.
 The residuals of all filters of a subsystem are 
 combined within one plot. The scatter is a measure of the
 error budget for single observations. The vertical dashed line 
 indicates the separation between low and high flux intervals.
 }
 \label{cmpfcsC}
\end{center}
\end{figure*}
 
We assess the overall consistency of the point source 
calibration in standard apertures by comparing the measured 
FCS in-band fluxes for all available heating powers with the
FCS power curves. The scatter is representative of the error
budget, except for any systematic biases in the SEDs of the
calibration sources. Each measurement passed through the 
standard signal processing chain and thus contains the same 
uncertainties as the science observations. 
The additional automatic transient fit we
applied during the calibration data processing is equivalent to
improvements the general observer can achieve through
manual processing by Interactive Analysis. We plot the ratio of
the in-band power taken from the calibration tables for the FCS 
(FCS power curves times FCS illumination tables) and the
measured FCS in-band powers. To indicate the flux regimes, we plot the
ratios versus in-band power on the detector. All data 
from a single
subinstrument appear together in one diagram. Figs~\ref{cmpfcsP}
and \ref{cmpfcsC} show the ratios derived for the bulk of
the mission for FCS\,1 and the ratios that
apply for FCS\,1/TRS\,2 before revolution 94. As
expected, the consistency improves with flux (i.e. the scatter
diminishes), because statistical
noise becomes less important and time constants of transients
decrease.

To obtain another quantitative representation, we considered the
deviations for individual filters and computed the maxima and 
the r.m.s. We also calculated the r.m.s. of the portions left
and right of the dotted lines in Figs \ref{cmpfcsP} and 
\ref{cmpfcsC}, to seek any differences between higher and
lower fluxes. Table~\ref{tabaccrt} contains the values
applicable for most of the mission while Table~\ref{tabaccpv}
shows the values for the time before rev.~94. These figures and
tables give an overview of the accuracy expected for an
observation of a point source in the standard aperture,
including measurements of background, source, FCS and cold
FCS. Since the FCS straylight measurement can in most cases be
replaced by the dark signal, the validity of this
assessment extends to even a larger number of observations. 
For multi-filter measurements, where an FCS measurement in only 
a single filter is available to determine $R$ (Eq.~\ref{resp_fcs}),
this FCS calibration can be transferred to a sky measurement in 
another filter of the same subsystem, because the ratios of 
transmissions between filters have been determined 
(see Sect.~\ref{ftof_sec}).
However, this transfer also accretes the uncertainty 
of the factor $\chi_{p,f}$ ($\approx 10$\,\%) for any filter 
different from the one that was actually used during the 
FCS measurement.

For a single staring observation performed in a standard 
aperture, we find scatters in the range
from 3 to 29\,\%, depending on detector, filter band and flux. 
The inhomogeneous distribution of data points suggests 
that systematic effects still dominate the error budget.
Since the data points are few and the properties of the
distribution, which may not be Gaussian, also depend on 
flux or signal level, the results must be interpreted with care. 
The average scatter (10\,\%) over all filters of detector P1 
appears to be the best metric for the accuracy that can 
be expected, but it is likely to be better.
A similar result applies to detector P2. 
Only P3 shows a much
larger scatter of about 20\,\% as a consequence of its degraded
performance in the radiation environment in space compared to the
laboratory (Lemke et al. \cite{lemke96}).
Better statistics are available for the 9 pixels of C100, that 
perform better than P3 although the detector material (Ge:Ga) 
is the same. We find an average scatter of 13\,\%. A better 
background subtraction, due to the raster mapping method we 
applied for the C-detectors, may also contribute to the 
reduced scatter. 
The results for the C200 detector also provide good statistics. 
The average scatter is about 10\,\%, fairly consistent over all 
filters.

These numbers serve as a guideline for the astronomer as to 
what accuracies to expect for a staring or raster 
observation that underwent a complete data reduction, involving 
all reduction steps
and all necessary supporting measurements, i.e. background,
FCS, FCS straylight. 
To assess the uncertainty of the FCS power curves alone, however, they 
only represent upper limits, since statistical uncertainties
decrease with the number of data points used. 

The uncertainties obtained are similar to or larger than the
accuracies quoted for the SEDs of the celestial standards.
This limits our ability to discriminate problematic 
standards in comparison to others. A first analysis
is given by Schulz (\cite{schulz2001}).

\subsection{Long-term Monitoring} \label{secmonitor}
In order to test the stability and the reproducibility of the
system, a set of non-variable celestial sources was monitored
at regular intervals throughout the mission. A bright and a
faint source were repeatedly measured in the standard filter band
of each detector, using the standard apertures in the case of the
P-detectors. For the bright source we used the planetary nebula
\object{NGC6543}, which was the prime photometric reference for
IRAS due to its good visibility and detectability from the MIR
to the FIR (Beichman et al. \cite{beichman88}). The faint
sources were chosen from the list of stellar calibration
standards with good visibility throughout the mission (see
Table~\ref{tabmon}). 

\begin{table}
\begin{center}
\begin{tabular}{llcccc} 
source  &  filter  &    apert. &  F$_{\nu}$ & F$_{\nu_\mathrm{meas.}}$ & $\sigma$   \\
        &          &           &    [Jy]    &  [Jy]     & [\%]       \\
\hline\noalign{\smallskip}
\object{HD172323} & P\_11.5  &     52$^{\prime\prime}$& 0.085 & 0.087 & 10.0  \\
\object{HR5986}   & P\_25    &     79$^{\prime\prime}$& 0.667 & 0.588 & 12.3  \\
\object{HR7310}   & P\_60    &    180$^{\prime\prime}$& 0.667 & 0.574 & 7.0   \\
\object{HR7310}   & C\_100   &     ~-                 & 0.229 & 0.250 & 10.6  \\
\object{HR6705}   & C\_160   &     ~-                 & 0.672 & 0.631 & 7.2   \\
\hline\noalign{\smallskip}
\object{NGC6543}  & P\_11.5  &     52$^{\prime\prime}$&   --  & 6.8   & 5.9   \\
\object{NGC6543}  & P\_25    &     79$^{\prime\prime}$&   --  & 107   & 1.8   \\
\object{NGC6543}  & P\_60    &    180$^{\prime\prime}$&   --  & 144   & 6.6   \\
\object{NGC6543}  & C\_100   &     ~-                 &   --  & 59.5  & 2.0   \\
\object{NGC6543}  & C\_160   &     ~-                 &   --  & 15.9  & 2.6   \\
\end{tabular}
\caption{\it Overview of results of the monitoring programme of 
faint and bright flux standards. The sources, filters and apertures used
appear in the first three columns; the fourth and fifth show
the flux densities calculated from model SEDs and the observed 
error-weighted mean values. Flux densities are not colour-corrected, 
i.e. they assume a $\nu F_{\nu}=const.$ spectrum. The last column shows
the standard deviation of the distribution of measurements.}
\label{tabmon}
\end{center}
\end{table}

The measurements on \object{NGC6543} were repeated every 5~weeks on
average; the fainter sources were measured every 2~weeks. The
planetary nebula was measured at the centre of each detector system,
after performing a background measurement at a position about 
5$\arcmin$ away. The faint sources were observed either in
nodding mode for the P-detectors or with a small raster of $3
\times 3$ or $4 \times 2$ for C100 and C200 respectively, to
minimize the uncertainties due to background subtraction.
We find the
reproducibility of measurements on bright sources to be around
2-3\,\%, except for P1 and P3 for which this is 6-7\,\%.
The fainter sources show larger scatter because uncertainties of
source, background, FCS and FCS straylight contribute equally to
the error, compared to the uncertainties only of source and FCS 
that contribute at higher fluxes. Photon noise and glitch 
noise also contribute relatively more at low fluxes. The 
reproducibility below 1~Jy ranges from 7 to 12\,\%.
A summary is given in Table~\ref{tabmon}.

\section{Summary} \label{summar}

We have described the photometric calibration of ISOPHOT using
celestial standards. We used point sources either stars, planets
or asteroids, to cover the full wavelength range. Standards at
all flux levels were used to calibrate nonlinearities.

Because the detector responsivity varied with time, stable
internal IR sources were calibrated against celestial standards
and measurements of
these references were included in each scientific observation.
The reproducibility of results derived from the ratios of these
signals was verified for high and low flux levels. We
corrected for non-ideal effects like nonlinearities,
imperfect optics and detector transients, and reduced
their influence on the measurements to a minimum.

The photometric calibration of the internal sources initiated
the definition of empirical FCS power curves and
FCS illumination tables as well as the introduction of matrices
of calibration factors, to account for pixel- and
filter-dependent attenuations, which are probably caused by
misalignments, diffraction and spatial gradients in
filter transmissions. 
A major step in achieving a consistent picture was the 
correction for the flux dependence of the responsivity.
A set of formulae describing the path from
raw detector signals to flux densities was derived, which
uses the detector responsivity to characterise the state of 
a detector pixel at the time of the observation.

To quantify the accuracies achievable by absolute
photometry, a comparison was made of the derived calibration curves of
the internal sources and the measured data.
Depending mainly on flux level, the achievable
uncertainties are around 10 to 20\,\%, but can exceed 30\,\%
under exceptional conditions and at low flux levels. The
filters of each subsystem were calibrated relative to each
other with accuracies better than 10\,\%. 

The use of point sources and standard apertures for each filter for
the aperture photometer defines a baseline within the
parameter space, which serves as a reference for the 
calibration of further instrument modes like chopped 
measurements, multi-aperture measurements and mapping.


\begin{acknowledgements}
We would like to thank U. Gr\"ozinger for vital support.
N. Lu and A. Wehrle are acknowledged for helpful comments
on an early version of the manuscript.
MC thanks the University of Florida for supporting 
his contributions to this work through subcontracts 
(under prime grants NAGW-4201 and NAG 5-3343) with VRI and,
in later years, through subcontract
UF99025 with Berkeley. 
\end{acknowledgements}


\end{document}